\documentclass[%
reprint,
twocolumn,
showpacs,preprintnumbers,
amsmath,amssymb,
aps,prl, 
nofootinbib,
longbibliography, 
floatfix,
]{revtex4-2}

\usepackage{units}
\usepackage{upgreek}
\usepackage{textcomp}
\usepackage{enumitem}  
\usepackage{graphicx}
\usepackage{epstopdf}
\usepackage{dcolumn}
\usepackage{bm}
\usepackage{hyperref}
\usepackage{booktabs}
\usepackage{gensymb}

\usepackage{xcolor}
\definecolor{mylinkcolor}{rgb}{0.0,0.0,0.66}
\hypersetup{colorlinks=true,linkcolor=mylinkcolor,urlcolor=mylinkcolor,citecolor=mylinkcolor,filecolor=mylinkcolor,breaklinks=true,linktocpage=true,linktoc=all}

\usepackage[normalem]{ulem}
\usepackage{color,soul} 
\usepackage{lipsum}  
\usepackage[export]{adjustbox}
\DeclareMathOperator{\arctanh}{arctanh}

\newcommand{\ie}{\textit{i.e.,}\ }

\newcommand{\sdist}{\kern 0.20em}
\renewcommand{\eqref}[1]{Eq.\sdist(\ref{#1})}

\mathchardef\mhyphen="2D

\usepackage{color}

\emergencystretch=\maxdimen
\begin{document}


\title{Ultra-strong Quantum Squeezing Mediated by Plasma Waves}
\author{Kenan Qu}
\affiliation{Department of Astrophysical Sciences, Princeton University,  Princeton, New Jersey 08544, USA \looseness=-1 }  
\author{Nathaniel J. Fisch}
\affiliation{Department of Astrophysical Sciences, Princeton University,  Princeton, New Jersey 08544, USA \looseness=-1 }

\date{\today}

\begin{abstract}
Quantum squeezed states enable precision measurements beyond the standard quantum limit, but conventional solid-state media fundamentally limit pump intensities to the ionization threshold. We demonstrate that plasma waves can mediate ultra-strong two-mode squeezing through stimulated Raman scattering, achieving up to ultrastrong squeezing using $\unit[10^{16}]{Wcm^{-2}}$ pump lasers. Employing two copropagating pump beams with frequency difference matching twice the plasma frequency, we generate quantum-correlated photon pairs through phonon-mediated four-wave mixing. The process exhibits remarkable thermal noise tolerance, allowing strong squeezing even with large thermal phonon numbers. This plasma-based approach produces squeezed states with ultrahigh photon numbers, opening new possibilities for strong-field applications across optical to X-ray wavelengths.
\end{abstract}


\maketitle

{\it Introduction.---}Quantum squeezed states~\cite{Walls_Sq1983, Agarwal_2012} constitute a foundational breakthrough in precision measurement, enabling quantum noise reduction below the shot-noise limit. This capability has proven essential for advancing the sensitivity of applications such as gravitational wave detection~\cite{np_LIGO_2013,PRX_LIGO_2023}, quantum imaging~\cite{Moodley_23}, and quantum spectroscopy~\cite{Mukamel_2020}. In addition, the nonclassical characteristics of squeezed states, namely a zero mean field and non-Poissonian photon statistics, have been exploited to substantially enhance the efficiency of high-harmonic generation~\cite{np_Even2023, Rasputnyi_np2024} and nonlinear Compton scattering~\cite{Khalaf_sciadv2023, ADP_KQ_2025}. Realizing these enhancements, however, demands access to squeezed light of extreme intensity or frequency, regimes that remain inaccessible with established generation schemes.


The production of squeezed light is presently circumscribed by the inherent material limitations of conventional solid-state media. These media, such as nonlinear crystals \cite{Franken_RMP1963, Canagasabey_09} and optical fibers \cite{Lin_06fiber, Garay_Palmett:23} are restricted to the optical-to-infrared spectral range and possess low damage thresholds, exhibiting thermal instability and undergoing ionization at pump intensities nearing $\unit[10^{12}\text{-}10^{14}]{Wcm^{-2}}$. This intensity threshold imposes a severe constraint on the attainable squeezing magnitude, with a record of $\unit[5.8]{dB}$ in a single-pass interaction~\cite{Kim_prl1994} and $\unit[10\text{-}15]{dB}$ if assisted with a cavity~\cite{PRL_15db_2016}. 
Breaking this intensity and frequency barrier requires moving to ionized plasma media, where pump powers can exceed solid-state limits by orders of magnitude. 

While quantum optical studies involving laser-plasma interactions have historically been limited~\cite{Milchberg_prl2008, Milchberg_pra2012, Milchberg_prl2014}, recent breakthroughs in attosecond physics have spurred a series of experiments demonstrating nonclassical light generation~\cite{PRA_Kominis2014, sc_HHG2016, nc_HHG2017, np_Lewenstein2021, PRL_Tsatrafyllis2019, GonoskovPRB2024, TzurPRR2024}. However, these approaches remain constrained by partial ionization~\cite{PRLStammer_2022, nc_HHG2020, PRXQuantum2023, TheidelPRX2024}, resulting in harmonic spectra that are not ideal for driving strong-field~\cite{np_Even2023, Rasputnyi_np2024, Khalaf_sciadv2023, ADP_KQ_2025} or quantum metrology~\cite{np_LIGO_2013, PRX_LIGO_2023, Moodley_23, Mukamel_2020} applications.
Fully ionized plasmas eliminate ionization constraints entirely, enabling pump intensities exceeding $\unit[10^{16}\text{-}10^{17}]{Wcm^{-2}}$ while supporting efficient nonlinear interactions~\cite{Shvets_1998, malkin99, Cheng2005, Ren_np2007, Ren_PoP2008, KQprl2017,  Vieux_Raman2017, np_Kirkwood2018}.
Our recent work~\cite{Qu_PRE_entangle24} demonstrated the feasibility of achieving $\unit[20]{dB}$ squeezing via plasma-based relativistic four-wave mixing (FWM)~\cite{Malkin_pre2020, Malkin_pre2020_2, Malkin_pre2022, Griffith2021}. Nevertheless, this $\chi^{(3)}$ process necessitates relativistic intensities and is susceptible to classical noise sources~\cite{Malkin2000, MinSup_np2023}.
We now present a fundamentally superior approach using stimulated Raman scattering (SRS)---a first-order process with dramatically higher growth rates that eliminates the primary classical noise sources while enabling unprecedented squeezing levels.


The key innovation employs phonon-mediated FWM, where plasma Langmuir waves serve as quantum intermediaries between optical modes. This mechanism produces quantum-correlated photon pairs through combined Stokes and anti-Stokes processes, generating ideal two-mode squeezed states with exceptional robustness to thermal noise. Specifically, two distinct SRS processes are involved: First, a pump photon converts into a photon at the Stokes sideband by creating a phonon, which is quantum-correlated with the emitted photon. However, because the phonon frequency is much lower than the optical frequency, this process alone does not yield squeezing. The crucial step is to replace the intermediate phonon with a second optical photon via the anti-Stokes process, where a pump photon is converted into an anti-Stokes photon by absorbing a phonon. This quantum state swapping effectively transfers the phonon’s quantum properties to the photon. By coherently combining these SRS pathways, the plasma mediates entanglement between the two optical output channels, producing a quantum two-mode squeezed state whose joint quadrature fluctuations are reduced below the shot-noise limit. Such a state serves as a key resource for continuous-variable quantum communication~\cite{CV_rmp2005}  and, more broadly, quantum information processing~\cite{GQI_rmp2012}. Importantly, this two-mode squeezed output can be readily converted into a single-mode squeezed state—highly desirable for quantum metrology and strong-field applications—by recombining the channels with a balanced beam splitter. 

\begin{figure}[h]
	\centering
	\includegraphics[width=\linewidth ]{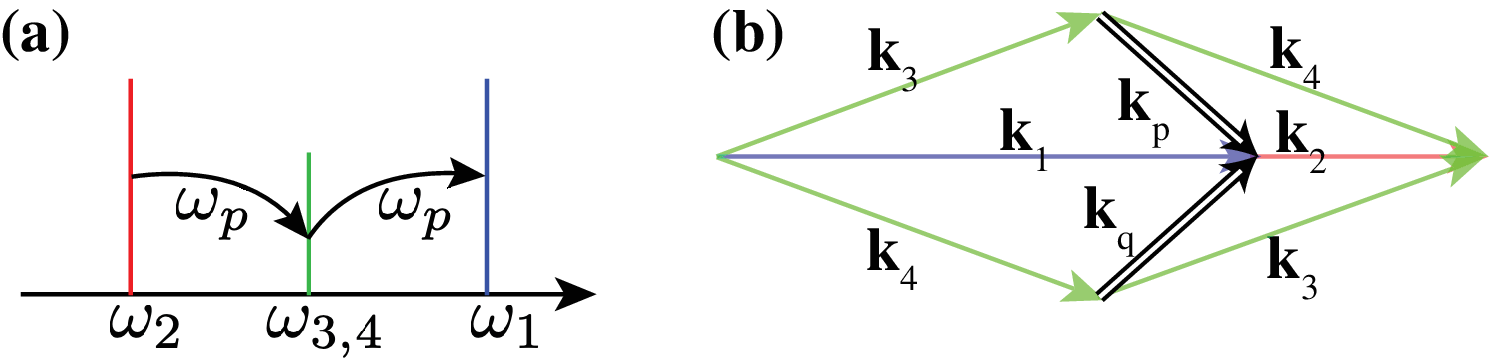} 
	\caption{(a) Frequency diagram of two pump lasers $\omega_{1,2}$ couples to the output mode $\omega_{3,4}$ through either emitting or absorbing a phonon $\omega_p$. (b) Wavevector diagram of the different laser and plasma waves that satisfy the phase matching condition.} 
	\label{Fig:diag}
\end{figure}

Similar SRS-mediated FWM processes have been experimentally realized in cold atoms~\cite{Rabl_prb2004, Parkins_prl2006} and the use of phonons was investigated in optomechanical platforms~\cite{Purdy_2013, Wangprl2013, Tianprl2013, KQnjp2014, KQpra2015, Magrini_2022}. In contrast, plasma enables access to unprecedented driving laser powers and flexible operating wavelengths, making it uniquely suited for ultra-strong squeezing across a broad range of spectra. Unlike relativistic FWM, where the laser couples directly to plasma electrons and the plasma wave contributes quantum noise via a separate channel~\cite{Balakin2003, Berger2004}, the present approach exploits the direct quantum mediation of the plasma wave, inherently suppressing classical noise sources. Because SRS is the fastest nonlinear process in laser-plasma interactions, it fundamentally enhances both the efficiency and noise resilience of squeezed light generation, allowing for the generation of squeezing magnitudes far beyond those achievable in solid-state or relativistic FWM schemes,

{\it Model and two-mode squeezing.---}The phonon-mediated four-wave mixing can be realized using two copropagating pump lasers with a frequency difference equal to the plasma frequency, as illustrated in Fig.~\ref{Fig:diag}(a). In the collective plasma regime $k\lambda_D <1$ ($k$ is the plasma phonon wavevector and $\lambda_D$ is the Debye length), each emission mode has a sharp spectrum~\cite{sheffield2010plasma}. Both pumps drive the interaction of a plasma phonon with emission photons at a degenerate frequency $\omega_{3,4}$, obeying the Manley-Rowe relations
\begin{equation}
\begin{gathered}
	\omega_1-\omega_{3,4} = \omega_{3,4} - \omega_2 = \omega_p, \\
	\bm{k}_1 - \bm{k}_3 = \bm{k}_4 - \bm{k}_2 = \bm{k}_p, \\
	\bm{k}_1 - \bm{k}_4 = \bm{k}_3 - \bm{k}_2 = \bm{k}_q,
\end{gathered}
\end{equation}
where $\omega_i$ and $\bm{k}_i$ are the frequency and wavevector for the electromagnetic wave $i=1,2,3,4$, which obey the dispersion relation $\omega_i^2 = c^2 k_i^2 + \omega_p^2$, and those of the plasma wave $i=p,q$. 
The interaction geometry, illustrated in Fig.~\ref{Fig:diag}(b), is essentially the inverse of the relativistic FWM~\cite{Malkin_pre2020}, in that it is the output pulses that cross at the small angle $\theta$ rather than the paraxial pump breams, with $\theta = \arccos[(k_1+k_2)/(2k_3)]$.  The issues of beam overlap are thus similarly encountered and addressed~\cite{Griffith2021}.   A key difference, of course, is that an additional resonance involving the plasma frequency is introduced, rather than avoided,  to mediate the SRS process.
Each of the plasma phonons uniquely couples a different Bogoliubov mode of the emitted photons, and their combined effect enables ideal two-mode squeezing. 

To understand the evolution of the quantum fluctuation of the emissions, we describe the phonon-mediated FWM with the Hamiltonian  $\hat{H} = \sum_{i=3,4} \omega_i \hat{a}_i^\dag \hat{a}_i + \omega_p \sum_{i=p,q} \hat{a}_i^\dag \hat{a}_i + \hat{H}_{int}$, where the interaction Hamiltonian is  
\begin{equation}
	\hat{H}_{int} = \hbar g (\alpha_1^*\hat{a}_3 \hat{p} + \alpha_2 \hat{a}_4^\dag \hat{p} + \alpha_1^*\hat{a}_4 \hat{q} + \alpha_2 \hat{a}_3^\dag \hat{q})  + h.c.. 
\end{equation}
Here, $\alpha_{1,2}$ represents the pump amplitude normalized to the laser vector potential $\bm{A}$ by $\bm\alpha_i = e\bm{A}_i/(m_ec^2)$, and $\hat{a}_{3,4}$, $\hat{p}$ and $\hat{q}$ are the annihilation operators for the emission field and the phonons, respectively. We adopt the standard normalization~\cite{Agarwal_2012} for the emission field $\bm{A}_i = \sum_{\bm{k}_i,s} \sqrt{\frac{\hbar}{2\omega_{\bm{k}_i,s_i}V\varepsilon_0}} \hat{a}_{\bm{k}_i,s} \hat{\sigma}_{\bm{k}_i,s} e^{-i\bm{k}\cdot\bm{r} - i\omega_i t} + h.c.$. The phonon mode is normalized~\cite{Qu_PRE_entangle24} to the plasma density fluctuation by $|\delta n_p|^2 \leftrightarrow \frac{\hbar e^2 k_p^2}{2V\epsilon_0 \omega_p^3m_e^2c^2} \hat p^\dag \hat p$ (and similarly for $\hat{q}$), which gives a coupling rate of $ g  = \frac{ck_p}{2}\sqrt\frac{\omega_p}{\omega_3}$. In these expressions, $e$ is the natural charge, $m_e$ being the electron rest mass, $V$ is the normalization volume, and $\hat\sigma_i$ is the polarization vector of the field $i$. The first two term of the Hamiltonian represent the self energy of the output modes and plasma wave modes. The last term $\hat{H}_{int}$ describes two types of interactions: the simultaneous creation of a photon-phonon pair ($\hat{a}_3$ and $\hat{p}$, or $\hat{a}_4$ and $\hat{q}$) and swapping between a photon and a phonon ($\hat{a}_3$ and $\hat{q}$, or $\hat{a}_4$ and $\hat{p}$). 

Since the phonon mode is coupled to both optical output channels, thermal noise increases the fluctuation in each optical mode. However, the noise injected into both optical channels is quantum-correlated, which makes it possible to isolate it using interference. This is most readily analyzed using a Bogoliubov transformation by defining two hybrid modes
\begin{equation} \label{eq:bogo}
	\hat{b}_3 = \hat{a}_3 \cosh r + \hat{a}_4^\dag \sinh r, \quad \hat{b}_4 = \hat{a}_4 \cosh r + \hat{a}_3^\dag \sinh r,
\end{equation}
where $r = \arctanh(|\alpha_1/\alpha_2|)$. Under this transformation, the variance of the joint-quadrature is related as $\langle X_a^2 \rangle = \langle X_b^2 \rangle e^{-2r}$ where $X_o = (\hat{o}_3 + \hat{o}_3^\dag + \hat{o}_4 + \hat{o}_4^\dag)/\sqrt2$ for $\hat{o} = \hat{a}$ or $\hat{b}$. 
Using an operator notation, the transform $\hat{b}_{3,4} = \hat{S}(r) \hat{a}_{3,4} \hat{S}^\dag(r)$ can be viewed as two-mode squeezing with $\hat{S}(r) = \exp[r(\hat{a}_3\hat{a}_4 - \hat{a}_3^\dag\hat{a}_4^\dag)]$. The joint vacuum of modes $\hat{b}_3$ and $\hat{b}_4$ is the two-mode squeezed vacuum state $|\xi\rangle = \hat{S}(r)|0,0\rangle$ with a mode occupancy $\langle \hat{b}_{3,4}^\dag \hat{b}_{3,4} \rangle = \langle \xi| \hat{a}_{3,4}^\dag \hat{a}_{3,4} |\xi\rangle = \sinh^2 r$. 
Thus, quantum squeezing can be achieved if at least one of the modes $\hat{b}_{3,4}$ is cooled to below $\sinh^2 r$. 

At $t=0$, the initial optical state is a vacuum state with a fluctuation $\langle X_a^2 \rangle = 1$, which implies $\langle X_b^2 \rangle = e^{2r}$. The initial plasma waves can be assumed to have an average thermal phonon number $\bar{n}_{th}$. 
Using the Bogoliubov modes we rewrite interaction Hamiltonian as 
\begin{equation}
	H_{int} = \hbar G(\hat{b}_4^\dag \hat{p} + \hat{b}_4 \hat{p}^\dag + \hat{b}_3^\dag \hat{q} + \hat{b}_3 \hat{q}^\dag), 
\end{equation}
where $G = \sqrt{\alpha_2^2-\alpha_1^2}g$. The Hamiltonian shows that each phonon mode $\hat{q}$, $\hat{p}$ is individually coupled to a Bogoliubov mode $\hat{b}_3$ or $\hat{b}_4$, respectively, allowing for the cooling of the hybrid modes. 

\begin{figure}[thb]
	\includegraphics[width=0.95\linewidth ]{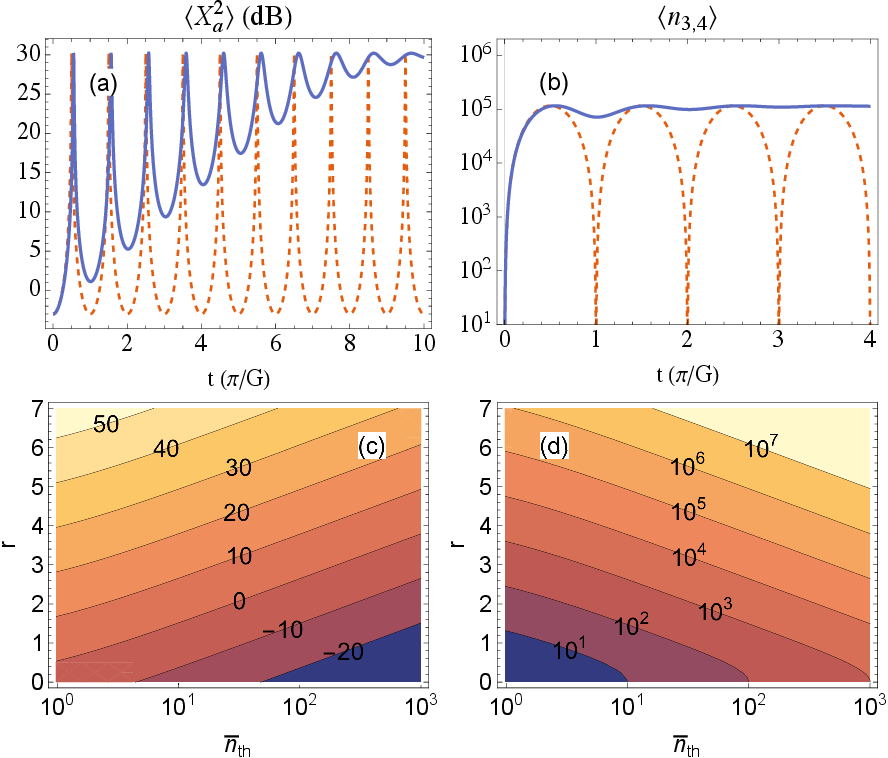}
	\caption{Evolution of (a) squeezing magnitude and (b) average output photon number versus time $t$. The thermal phonon number $\bar{n}_{th}=10$. The plasma wave decay rate $\kappa/G=0$ (dashed) and $0.3$ (solid). The squeezing magnitude in dB (c) and average output photon number (d) in the asymptotic limit for different values of $r$ and $\bar{n}_{th}$. } 
	\label{fig:squ}
\end{figure}

In the simplest model with zero plasma wave damping, the photon and phonon occupancy are exchanged with a period of $\pi/G$. 
At time $t=(n+\frac12)\pi/G$ for integer $n$'s, the Bogoliubov mode would have a variance $\langle X_b^2 \rangle = 2\bar{n}_{th} +1$. The covariance of the two optical output becomes $\langle X_a^2 \rangle = (2\bar{n}_{th} +1)e^{-2r}$ which indicates quantum squeezing if this value is less than the covariance of a joint-vacuum mode. The dynamics of covariance is plotted as a dashed curve in Fig.~\ref{fig:squ}(a). 
Because  $r$ can be controlled by the asymmetry of the pump amplitudes, \ie $\tanh r = |\alpha_1/\alpha_2|$, the output can be squeezed for an arbitrary thermal phonon number $\bar{n}_{th}$ as long as $(2\bar{n}_{th} +1) < e^{2r}$, as shown in Fig.~\ref{fig:squ}(c).  Notably, the average photon number [dashed curve in Fig.~\ref{fig:squ}(b)] grows at each of the output mode $\hat{a}_{3,4}$ even as their covariance asymptotically decreases. Note, however, a large values of $\bar{n}_{th}$ (and $r$) requires asymptotically equal pump amplitudes ($|\alpha_1| \sim |\alpha_2|$), which reduces the coupling rate $G$ and increases the mode conversion time. Therefore, a large thermal phonon number $\bar{n}_{th}$ can only be overcome at the cost of a reduced ``effective'' coupling rate and, consequently, a longer laser-plasma interaction region.

This result is significant, as it provides a pathway to produce a quantum two-mode squeezed state through two Raman scattering processes, even when the mediating plasma waves have thermal fluctuations. As shown in Fig.~\ref{fig:squ}(c) and (d), even with a thermal phonon number of $\bar{n}_{th}=100$, one can achieve a squeezing magnitude of $\unit[40]{dB}$ if choosing $r=7$, with the output photon number in each mode reaching  approximately $\sim10^{18}$.  This immunity to thermal phonons is possible for two reasons. First, the phonon-mediated FWM produces quantum-entangled photon pairs, which minimizes the quantum fluctuation between the two output modes. Second, because the Bogoliubov mode and the phonon mode are eigenmodes of the laser-plasma system, noise from the thermal phonon modes is periodically decoupled from the optical output, allowing for noise isolation and a large magnitude of quantum squeezing. 

Interestingly, the thermal phonon number $\bar{n}_{th} = k_BT/(\hbar\omega_p)$ is inversely proportional to the mode frequency $\omega_p$,  where $k_B$ is the Boltzmann constant, and $T$ is the temperature. Due to their  extremely high eigenfrequency $\omega_p$, plasma waves benefit from a relatively low thermal phonon number compared to solid-state crystals at room temperature. While thermal phonons causing inelastic Brillouin scattering in crystals~\cite{CrystalPhonon_pra2009} have frequencies in the GHz range, plasma frequencies for plasma photonics are typically in the tens or even hundreds of THz.  Even if a plasma have $\unit[10]{eV}$ or $\unit[10^6]{K}$ temperature, its phonon number is still below that of a nonlinear crystal at room temperature.

{\it Plasma wave damping.---}In plasmas, finite temperature not only causes incoherent thermal phonons, but also damping of the plasma waves. Quantum mechanically, plasma wave damping is treated by coupling the phonon modes to a constant noise reservoir. In the Heisenberg picture, the operators evolve in the rotating frame according to the quantum Langevin equations 
$\partial_t \hat{b}_3 = iG \hat{q}$, $\partial_t \hat{q} = -\kappa \hat{q} + iG \hat{b}_3 - \sqrt{2\kappa} \hat{q}_{in}$, $\partial_t \hat{b}_4 = iG \hat{p}$, and $\partial_t \hat{p} = -\kappa \hat{p} + iG \hat{b}_4 - \sqrt{2\kappa} \hat{p}_{in}$, 
where $\kappa = \sqrt{\pi/8}[\omega_p(k_{p,q}\lambda_D)^{-3}]\exp[-\frac12(k_{p,q}\lambda_D)^{-2}]$ is the plasma wave damping rate dominated by Landau damping under the parameters of interest, and $\lambda_D = \sqrt{k_BT/(m_e\omega_p^2)}$ is the plasma Debye length. 
The noise input $\hat{p}_{in}$ and $\hat{p}_{in}$ obey the correlation function~\cite{Lax_pr1966}
\begin{align}\label{eq:corr}
	\langle \hat{p}_{in}^\dag(t) \hat{p}_{in}(t') \rangle &=  \langle \hat{q}_{in}^\dag(t) \hat{q}_{in}(t') \rangle = \bar{n}_{th} \delta(t-t'), \\
	\langle \hat{p}_{in}(t) \hat{p}_{in}^\dag(t') \rangle &= \langle \hat{q}_{in}(t) \hat{q}_{in}^\dag(t') \rangle = (\bar{n}_{th}+1) \delta(t-t'). 
\end{align}
The quantum Langevin equations show two uncoupled decaying oscillation of the Bogoliubov modes  
\begin{align}
	\hat{b}_3(t) &= e^{-\frac{\kappa}{2}t} \frac{G}{\Delta} \left[ \cos(\Delta t - \varphi)\hat{b}_3(0)   + \sin(\Delta t) \hat{q}(0) \right]   \nonumber \\
	&\quad - \sqrt{2\kappa} \int_0^t e^{-\frac{\kappa}{2}\tau} \frac{G}{\Delta} \sin(\Delta\tau) \hat{q}_{in}(\tau) d\tau, \\
	\hat{b}_4(t) &= e^{-\frac{\kappa}{2}t} \frac{G}{\Delta} \left[ \cos(\Delta t - \varphi)\hat{b}_4(0)  + \sin(\Delta t) \hat{p}(0) \right]    \nonumber \\
	&\quad - \sqrt{2\kappa} \int_0^t e^{-\frac{\kappa}{2}\tau} \frac{G}{\Delta} \sin(\Delta\tau) \hat{p}_{in}(\tau) d\tau, 
\end{align}
where $\Delta = \sqrt{G^2-(\kappa/2)^2}$ and $\tan\varphi = \kappa/(2\Delta)$. The operator $\hat{a}_{3,4}$ for the output modes can be obtained using Eq.~(\ref{eq:bogo}). One finds that the covariance of quadrature $X_a = (\hat{a}_3 + \hat{a}_3^\dag + \hat{a}_4 + \hat{a}_4^\dag)/\sqrt2$ and the average output photon number $\langle n_{3,4} \rangle \equiv \langle \hat{a}_{3,4}^\dag \hat{a}_{3,4} \rangle$ are
\begin{align}
	\langle X_a^2 \rangle &= \mathcal{M} + (1-\mathcal{M}) (\bar{n}_{th} + 1/2) e^{-2r}, \\
	\langle n_{3,4} \rangle &= (1-\mathcal{M}) [\bar{n}_{th} \cosh^2r + (\bar{n}_{th}+1) \sinh^2r], \label{eq:10}
\end{align}
where $\mathcal{M} = e^{-\kappa t}(G/\Delta )^2 \cos^2(\Delta t-\varphi)$ represents the contribution from the vacuum field. The term proportional to $1-\mathcal{M}$ signifies the contribution from thermal phonons. To illustrate the effect of finite plasma wave damping, we plot the  evolution of output modes with $\kappa/G=0.3$ as solid curves in Fig.~\ref{fig:squ}(a) and (b). Intriguingly, plasma wave damping does not diminish the squeezing magnitude or the output photon number. On the contrary, a finite plasma wave damping rate $\kappa$ allows stable squeezing  of the vacuum fluctuation in the asymptotic limit $t \to \infty$. In this limit, the covariance $\langle X_a^2 \rangle$ is completely determined by the engineered noise reservoir $(\bar{n}_{th} + 1/2) e^{-2r}$. 
The output photon number in each mode also quickly approaches its maximum value and remains constant without oscillation.

{\it Single-mode squeezing output.---}A two-mode squeezed state exhibits quantum noise reduction below the shot-noise level in a linear combination of the quadratures of the two fields. By performing an inverse operation---combining both outputs using a balanced beam splitter---this state can be converted into a single-mode squeezed state. Such states can directly drive efficient higher harmonic generation and nonlinear Compton scattering~\cite{np_Even2023, Rasputnyi_np2024, Khalaf_sciadv2023, ADP_KQ_2025}, and they are particularly useful in applications demanding minimal field fluctuation~\cite{np_LIGO_2013,PRX_LIGO_2023} to achieve a high signal-to-noise ratio. Using a phase shifter and a balanced beam splitter, the output fields are converted into $\hat{c}_1 = (\hat{a}_3 + \hat{a}_4)/\sqrt2$ and $\hat{c}_2 = i(\hat{a}_3 - \hat{a}_4)/\sqrt2$ on each side. The $X_{c1}$ quadrature of the $\hat{c}_1$ mode is then $X_{c1} = (\hat{c}_1 + \hat{c}_1^\dag)/\sqrt2 = (\hat{a}_3 + \hat{a}_3^\dag + \hat{a}_4 + \hat{a}_4^\dag)/2$. It is identical to the covariance $X_a$ of the two-mode squeezed state, which has shown to exhibit reduced fluctuation. The complementary quadrature $Y_{c1} = (\hat{c}_1 - \hat{c}_1^\dag)/(\sqrt2i) = (\hat{a}_3 - \hat{a}_3^\dag + \hat{a}_4 - \hat{a}_4^\dag)/(2i)$ can be shown to have antisqueezed fluctuations.

{\it Experimental implementation.---}A two-mode squeezed output could be obtained at an angle $\theta$ if two copropagating pump pulses are directed into a plasma with their frequency detuning precisely matching twice the plasma frequency $\omega_p$. A classical detection yields equal intensity on both modes. The magnitude of quantum squeezing can be measured using homodyne detection~\cite{Walls_Sq1983, Agarwal_2012} after the outputs are converted into the aforementioned single-mode squeezing states.  Importantly, both for optimization and for elucidating the process, the angle $\theta$ is adjustable through the plasma density. 
 
Several factors need to be taken into consideration for achieving the maximum squeezing magnitude. The first factor is the pump amplitude $\alpha_{1,2}$ and their ratio $\alpha_1/\alpha_2$. In general, the mode coupling rate scales linearly with pump amplitudes but stable SRS requires $\alpha_{1,2}<1$ to avoid  relativistic effects~\cite{Umstadter_2003}. Their ratio should be controlled such that $\alpha_1/\alpha_2\approx 1$ to achieve the strongest squeezing, but a value too near unity suppresses the ``effective'' growth rate $G=\sqrt{\alpha_2^2 - \alpha_1^2}g$. 
For given plasma length $L$, one should first determine the value of $G_\mathrm{opt} = \pi v_{g3}/(2L)$ such that $\pi/2$ growth length is achieved. 

The second factor is the ratio of plasma and laser frequencies $\omega_p/\omega_3$. A larger ratio benefits the SRS photon scattering rate ${g\approx (\omega_p/2)\sqrt{\omega_p/\omega_3}}$, but it also increases the slippage the two pump pulses due to different group velocities ${v_{gi} = c\sqrt{1-(\omega_p/\omega_i)^2}}$. For given laser pulse durations, the slippage could limit the interaction distance as ${v_{g1}-v_{g2} \approx 3(\omega_p/\omega_3)^2c}$ for ${\omega_p/\omega_3\ll 1}$, which increases faster than $g$. Assuming a pulse duration $\tau$ and an interaction time equal to the pump slippage time $t_\mathrm{int} = v_{g3}\tau/(v_{g1}-v_{g2}) \approx 3(\omega_p/\omega_3)^2\tau$, a $\pi/2$ growth length requires 
\begin{equation}
	G t_\mathrm{int} \approx \frac16\alpha_1\sqrt{2\delta}   \left(\frac{\omega_3}{\omega_p}\right)^\frac32 \omega_3\tau = \frac\pi2
\end{equation}
where ${\delta=\alpha_2/\alpha_1 -1}$ denotes the difference between the pump amplitudes. 
A larger plasma frequency also demands a large pump detuning and increases the emission angle $\theta = \arccos[(k_1+k_2)/(2k_3)]$, which may require a plasma channel~\cite{York_prl2008} to confine the emitted photons and achieve a sufficient interaction distance. Such a channel might also offset the pump slippage as a recent study~\cite{palastro2025} shows its role in controlling the pulse group velocity similar to an optical fiber.
Slippage of the outputs, however, is not a concern, because ${v_{3,4}\cos\theta > v_2}$, \ie they remain larger than the group velocity of the lower-frequency pump pulse even when projected onto the pump direction .

The third factor is the plasma temperature $T$, which determines both the thermal phonon number $\bar{n}_{th}$ and plasma wave damping rate $\kappa$. A lower plasma temperature advantageously yields lower classical noise. But a finite temperature, and thus a finite $\kappa$, could be beneficial by  damping the oscillatory behavior of the squeezing magnitude over an extended interaction length. Combined with the value of $\alpha_1/\alpha_2=\tanh r$, the theoretical maximum squeezing magnitude can be found, \ie $(2\bar{n}_{th}+1)/e^{2r}$. 


As an example, consider a set of moderate experimental parameters: a mildly relativistic two-color pump laser with amplitudes $\alpha_{1,2} \approx 0.1$ (intensities near $\unit[10^{16}]{Wcm^{-2}}$) and wavelengths $\unit[0.8]{\mu m}$ and $\unit[0.89]{\mu m}$, and a plasma with a density $\unit[3.9\times10^{18}]{cm^{-3}}$ ($1/400$ the critical density). This allows for a long laser-plasma interaction distance without inducing parasitic instabilities~\cite{PBA_PRX2019}. The slippage distance for pump pulse duration of $\unit[100]{fs}$ pulse is about $\unit[10]{cm}$, but we choose a $\unit[1]{cm}$ interaction distance where the peaks of the Gaussian pulses nearly overlap. Reaching $\pi/2$ growth distance thus requires a growth rate $G=2.13\times10^{-5}\omega_1$ and the pump amplitudes differ by $\delta = 6\times10^{-4}$ (correspondingly $r=4$).  For a cold plasma with $T=\unit[1]{eV}$, \ie $\bar{n}_{th}\approx 10$, the maximum squeezing magnitude can achieve near $\unit[20]{dB}$ below the vacuum level despite an average photon number of over $1000$ in each mode.The output photons are emitted at an angle of $0.16^\circ$. While its longitudinal effect on the group velocity slippage with the pump is negligible, the transverse divergence after propagating for $\unit[1]{cm}$ distance could reach $\unit[28]{\mu m}$, which is within the spot size of a moderately intense laser.

If a plasma channel~\cite{York_prl2008} is used to confine the emitted photons, the interaction distance or the plasma density can be substantially increased. If we increase the plasma length to $\unit[2.5]{cm}$ and its density to $1/100$ the critical density, the required growth rate reduces to $G=9\times10^{-6}\omega_1$ and $\delta = 1.2\times10^{-5}$ (correspondingly $r=6$).  With plasma thermal phonon number $\bar{n}_{th}\approx 10$, the maximum squeezing magnitude can acheive $\unit[40]{dB}$ below the vacuum level with each output mode containing near $10^{6}$ photons.


Plasma SRS operates across unprecedented spectral ranges from optical to X-ray wavelengths. The interaction rate scales as $(\omega_p^3/\omega_3)^\frac12$, making shorter wavelengths increasingly favorable. Experimental demonstrations of SRS already exist in UV/EUV~\cite{Tanaka_prl1982, Wilson_pre2012} and even the X-ray regime~\cite{Manes_1985, Drake_pra1989, Glenzer_prl2003, Glenzer_prl2007}. X-ray squeezing becomes particularly attractive: wavelengths near $\unit[1]{nm}$ and plasma density near $\unit[10^{21}]{cm^{-3}}$ can achieve $\unit[40]{dB}$ squeezing with pump parameters ($\alpha_1 \tau$) three orders of magnitude smaller than optical examples. This scaling advantage positions plasma photonics as the optimal platform for short-wavelength quantum applications.

\begin{acknowledgments}
	This work was supported by  NNSA Grant No. DE-NA0004167. 
\end{acknowledgments}  

\bibliography{entangle}

\begin{thebibliography}{70}%
\makeatletter
\providecommand \@ifxundefined [1]{%
 \@ifx{#1\undefined}
}%
\providecommand \@ifnum [1]{%
 \ifnum #1\expandafter \@firstoftwo
 \else \expandafter \@secondoftwo
 \fi
}%
\providecommand \@ifx [1]{%
 \ifx #1\expandafter \@firstoftwo
 \else \expandafter \@secondoftwo
 \fi
}%
\providecommand \natexlab [1]{#1}%
\providecommand \enquote  [1]{``#1''}%
\providecommand \bibnamefont  [1]{#1}%
\providecommand \bibfnamefont [1]{#1}%
\providecommand \citenamefont [1]{#1}%
\providecommand \href@noop [0]{\@secondoftwo}%
\providecommand \href [0]{\begingroup \@sanitize@url \@href}%
\providecommand \@href[1]{\@@startlink{#1}\@@href}%
\providecommand \@@href[1]{\endgroup#1\@@endlink}%
\providecommand \@sanitize@url [0]{\catcode `\\12\catcode `\$12\catcode
  `\&12\catcode `\#12\catcode `\^12\catcode `\_12\catcode `\%12\relax}%
\providecommand \@@startlink[1]{}%
\providecommand \@@endlink[0]{}%
\providecommand \url  [0]{\begingroup\@sanitize@url \@url }%
\providecommand \@url [1]{\endgroup\@href {#1}{\urlprefix }}%
\providecommand \urlprefix  [0]{URL }%
\providecommand \Eprint [0]{\href }%
\providecommand \doibase [0]{https://doi.org/}%
\providecommand \selectlanguage [0]{\@gobble}%
\providecommand \bibinfo  [0]{\@secondoftwo}%
\providecommand \bibfield  [0]{\@secondoftwo}%
\providecommand \translation [1]{[#1]}%
\providecommand \BibitemOpen [0]{}%
\providecommand \bibitemStop [0]{}%
\providecommand \bibitemNoStop [0]{.\EOS\space}%
\providecommand \EOS [0]{\spacefactor3000\relax}%
\providecommand \BibitemShut  [1]{\csname bibitem#1\endcsname}%
\let\auto@bib@innerbib\@empty
\bibitem [{\citenamefont {Walls}(1983)}]{Walls_Sq1983}%
  \BibitemOpen
  \bibfield  {author} {\bibinfo {author} {\bibfnamefont {D.~F.}\ \bibnamefont
  {Walls}},\ }\bibfield  {title} {\bibinfo {title} {Squeezed states of light},\
  }\href {https://doi.org/10.1038/306141a0} {\bibfield  {journal} {\bibinfo
  {journal} {Nature}\ }\textbf {\bibinfo {volume} {306}},\ \bibinfo {pages}
  {141} (\bibinfo {year} {1983})}\BibitemShut {NoStop}%
\bibitem [{\citenamefont {Agarwal}(2012)}]{Agarwal_2012}%
  \BibitemOpen
  \bibfield  {author} {\bibinfo {author} {\bibfnamefont {G.~S.}\ \bibnamefont
  {Agarwal}},\ }\href@noop {} {\emph {\bibinfo {title} {Quantum Optics}}}\
  (\bibinfo  {publisher} {Cambridge University Press},\ \bibinfo {year}
  {2012})\BibitemShut {NoStop}%
\bibitem [{\citenamefont {Aasi}\ \emph {et~al.}(2013)\citenamefont {Aasi} \emph
  {et~al.}}]{np_LIGO_2013}%
  \BibitemOpen
  \bibfield  {author} {\bibinfo {author} {\bibfnamefont {J.}~\bibnamefont
  {Aasi}} \emph {et~al.},\ }\bibfield  {title} {\bibinfo {title} {Enhanced
  sensitivity of the ligo gravitational wave detector by using squeezed states
  of light},\ }\href {https://doi.org/10.1038/nphoton.2013.177} {\bibfield
  {journal} {\bibinfo  {journal} {Nature Photonics}\ }\textbf {\bibinfo
  {volume} {7}},\ \bibinfo {pages} {613} (\bibinfo {year} {2013})}\BibitemShut
  {NoStop}%
\bibitem [{\citenamefont {Ganapathy}\ \emph {et~al.}(2023)\citenamefont
  {Ganapathy} \emph {et~al.}}]{PRX_LIGO_2023}%
  \BibitemOpen
  \bibfield  {author} {\bibinfo {author} {\bibfnamefont {D.}~\bibnamefont
  {Ganapathy}} \emph {et~al.} (\bibinfo {collaboration} {LIGO O4 Detector
  Collaboration}),\ }\bibfield  {title} {\bibinfo {title} {Broadband quantum
  enhancement of the ligo detectors with frequency-dependent squeezing},\
  }\href {https://doi.org/10.1103/PhysRevX.13.041021} {\bibfield  {journal}
  {\bibinfo  {journal} {Phys. Rev. X}\ }\textbf {\bibinfo {volume} {13}},\
  \bibinfo {pages} {041021} (\bibinfo {year} {2023})}\BibitemShut {NoStop}%
\bibitem [{\citenamefont {Moodley}\ and\ \citenamefont
  {Forbes}(2023)}]{Moodley_23}%
  \BibitemOpen
  \bibfield  {author} {\bibinfo {author} {\bibfnamefont {C.}~\bibnamefont
  {Moodley}}\ and\ \bibinfo {author} {\bibfnamefont {A.}~\bibnamefont
  {Forbes}},\ }\bibfield  {title} {\bibinfo {title} {All-digital quantum ghost
  imaging: tutorial},\ }\href {https://doi.org/10.1364/JOSAB.489100} {\bibfield
   {journal} {\bibinfo  {journal} {J. Opt. Soc. Am. B}\ }\textbf {\bibinfo
  {volume} {40}},\ \bibinfo {pages} {3073} (\bibinfo {year}
  {2023})}\BibitemShut {NoStop}%
\bibitem [{\citenamefont {Mukamel}\ \emph {et~al.}(2020)\citenamefont
  {Mukamel}, \citenamefont {Freyberger}, \citenamefont {Schleich},
  \citenamefont {Bellini}, \citenamefont {Zavatta}, \citenamefont {Leuchs},
  \citenamefont {Silberhorn}, \citenamefont {Boyd}, \citenamefont
  {Sánchez-Soto}, \citenamefont {Stefanov}, \citenamefont {Barbieri},
  \citenamefont {Paterova}, \citenamefont {Krivitsky}, \citenamefont {Shwartz},
  \citenamefont {Tamasaku}, \citenamefont {Dorfman}, \citenamefont {Schlawin},
  \citenamefont {Sandoghdar}, \citenamefont {Raymer}, \citenamefont {Marcus},
  \citenamefont {Varnavski}, \citenamefont {Goodson}, \citenamefont {Zhou},
  \citenamefont {Shi}, \citenamefont {Asban}, \citenamefont {Scully},
  \citenamefont {Agarwal}, \citenamefont {Peng}, \citenamefont {Sokolov},
  \citenamefont {Zhang}, \citenamefont {Zubairy}, \citenamefont {Vartanyants},
  \citenamefont {del Valle},\ and\ \citenamefont {Laussy}}]{Mukamel_2020}%
  \BibitemOpen
  \bibfield  {author} {\bibinfo {author} {\bibfnamefont {S.}~\bibnamefont
  {Mukamel}}, \bibinfo {author} {\bibfnamefont {M.}~\bibnamefont {Freyberger}},
  \bibinfo {author} {\bibfnamefont {W.}~\bibnamefont {Schleich}}, \bibinfo
  {author} {\bibfnamefont {M.}~\bibnamefont {Bellini}}, \bibinfo {author}
  {\bibfnamefont {A.}~\bibnamefont {Zavatta}}, \bibinfo {author} {\bibfnamefont
  {G.}~\bibnamefont {Leuchs}}, \bibinfo {author} {\bibfnamefont
  {C.}~\bibnamefont {Silberhorn}}, \bibinfo {author} {\bibfnamefont {R.~W.}\
  \bibnamefont {Boyd}}, \bibinfo {author} {\bibfnamefont {L.~L.}\ \bibnamefont
  {Sánchez-Soto}}, \bibinfo {author} {\bibfnamefont {A.}~\bibnamefont
  {Stefanov}}, \bibinfo {author} {\bibfnamefont {M.}~\bibnamefont {Barbieri}},
  \bibinfo {author} {\bibfnamefont {A.}~\bibnamefont {Paterova}}, \bibinfo
  {author} {\bibfnamefont {L.}~\bibnamefont {Krivitsky}}, \bibinfo {author}
  {\bibfnamefont {S.}~\bibnamefont {Shwartz}}, \bibinfo {author} {\bibfnamefont
  {K.}~\bibnamefont {Tamasaku}}, \bibinfo {author} {\bibfnamefont
  {K.}~\bibnamefont {Dorfman}}, \bibinfo {author} {\bibfnamefont
  {F.}~\bibnamefont {Schlawin}}, \bibinfo {author} {\bibfnamefont
  {V.}~\bibnamefont {Sandoghdar}}, \bibinfo {author} {\bibfnamefont
  {M.}~\bibnamefont {Raymer}}, \bibinfo {author} {\bibfnamefont
  {A.}~\bibnamefont {Marcus}}, \bibinfo {author} {\bibfnamefont
  {O.}~\bibnamefont {Varnavski}}, \bibinfo {author} {\bibfnamefont
  {T.}~\bibnamefont {Goodson}}, \bibinfo {author} {\bibfnamefont {Z.-Y.}\
  \bibnamefont {Zhou}}, \bibinfo {author} {\bibfnamefont {B.-S.}\ \bibnamefont
  {Shi}}, \bibinfo {author} {\bibfnamefont {S.}~\bibnamefont {Asban}}, \bibinfo
  {author} {\bibfnamefont {M.}~\bibnamefont {Scully}}, \bibinfo {author}
  {\bibfnamefont {G.}~\bibnamefont {Agarwal}}, \bibinfo {author} {\bibfnamefont
  {T.}~\bibnamefont {Peng}}, \bibinfo {author} {\bibfnamefont {A.~V.}\
  \bibnamefont {Sokolov}}, \bibinfo {author} {\bibfnamefont {Z.-D.}\
  \bibnamefont {Zhang}}, \bibinfo {author} {\bibfnamefont {M.~S.}\ \bibnamefont
  {Zubairy}}, \bibinfo {author} {\bibfnamefont {I.~A.}\ \bibnamefont
  {Vartanyants}}, \bibinfo {author} {\bibfnamefont {E.}~\bibnamefont {del
  Valle}},\ and\ \bibinfo {author} {\bibfnamefont {F.}~\bibnamefont {Laussy}},\
  }\bibfield  {title} {\bibinfo {title} {Roadmap on quantum light
  spectroscopy},\ }\href {https://doi.org/10.1088/1361-6455/ab69a8} {\bibfield
  {journal} {\bibinfo  {journal} {Journal of Physics B: Atomic, Molecular and
  Optical Physics}\ }\textbf {\bibinfo {volume} {53}},\ \bibinfo {pages}
  {072002} (\bibinfo {year} {2020})}\BibitemShut {NoStop}%
\bibitem [{\citenamefont {Even~Tzur}\ \emph {et~al.}(2023)\citenamefont
  {Even~Tzur}, \citenamefont {Birk}, \citenamefont {Gorlach}, \citenamefont
  {Kr{\"u}ger}, \citenamefont {Kaminer},\ and\ \citenamefont
  {Cohen}}]{np_Even2023}%
  \BibitemOpen
  \bibfield  {author} {\bibinfo {author} {\bibfnamefont {M.}~\bibnamefont
  {Even~Tzur}}, \bibinfo {author} {\bibfnamefont {M.}~\bibnamefont {Birk}},
  \bibinfo {author} {\bibfnamefont {A.}~\bibnamefont {Gorlach}}, \bibinfo
  {author} {\bibfnamefont {M.}~\bibnamefont {Kr{\"u}ger}}, \bibinfo {author}
  {\bibfnamefont {I.}~\bibnamefont {Kaminer}},\ and\ \bibinfo {author}
  {\bibfnamefont {O.}~\bibnamefont {Cohen}},\ }\bibfield  {title} {\bibinfo
  {title} {Photon-statistics force in ultrafast electron dynamics},\ }\href
  {https://doi.org/10.1038/s41566-023-01209-w} {\bibfield  {journal} {\bibinfo
  {journal} {Nature Photonics}\ }\textbf {\bibinfo {volume} {17}},\ \bibinfo
  {pages} {501} (\bibinfo {year} {2023})}\BibitemShut {NoStop}%
\bibitem [{\citenamefont {Rasputnyi}\ \emph {et~al.}(2024)\citenamefont
  {Rasputnyi}, \citenamefont {Chen}, \citenamefont {Birk}, \citenamefont
  {Cohen}, \citenamefont {Kaminer}, \citenamefont {Kr{\"u}ger}, \citenamefont
  {Seletskiy}, \citenamefont {Chekhova},\ and\ \citenamefont
  {Tani}}]{Rasputnyi_np2024}%
  \BibitemOpen
  \bibfield  {author} {\bibinfo {author} {\bibfnamefont {A.}~\bibnamefont
  {Rasputnyi}}, \bibinfo {author} {\bibfnamefont {Z.}~\bibnamefont {Chen}},
  \bibinfo {author} {\bibfnamefont {M.}~\bibnamefont {Birk}}, \bibinfo {author}
  {\bibfnamefont {O.}~\bibnamefont {Cohen}}, \bibinfo {author} {\bibfnamefont
  {I.}~\bibnamefont {Kaminer}}, \bibinfo {author} {\bibfnamefont
  {M.}~\bibnamefont {Kr{\"u}ger}}, \bibinfo {author} {\bibfnamefont
  {D.}~\bibnamefont {Seletskiy}}, \bibinfo {author} {\bibfnamefont
  {M.}~\bibnamefont {Chekhova}},\ and\ \bibinfo {author} {\bibfnamefont
  {F.}~\bibnamefont {Tani}},\ }\bibfield  {title} {\bibinfo {title}
  {High-harmonic generation by a bright squeezed vacuum},\ }\href
  {https://doi.org/10.1038/s41567-024-02659-x} {\bibfield  {journal} {\bibinfo
  {journal} {Nature Physics}\ }\textbf {\bibinfo {volume} {20}},\ \bibinfo
  {pages} {1960} (\bibinfo {year} {2024})}\BibitemShut {NoStop}%
\bibitem [{\citenamefont {Khalaf}\ and\ \citenamefont
  {Kaminer}(2023)}]{Khalaf_sciadv2023}%
  \BibitemOpen
  \bibfield  {author} {\bibinfo {author} {\bibfnamefont {M.}~\bibnamefont
  {Khalaf}}\ and\ \bibinfo {author} {\bibfnamefont {I.}~\bibnamefont
  {Kaminer}},\ }\bibfield  {title} {\bibinfo {title} {Compton scattering driven
  by intense quantum light},\ }\href {https://doi.org/10.1126/sciadv.ade0932}
  {\bibfield  {journal} {\bibinfo  {journal} {Science Advances}\ }\textbf
  {\bibinfo {volume} {9}},\ \bibinfo {pages} {eade0932} (\bibinfo {year}
  {2023})}\BibitemShut {NoStop}%
\bibitem [{\citenamefont {Di~Piazza}\ and\ \citenamefont
  {Qu}(2025)}]{ADP_KQ_2025}%
  \BibitemOpen
  \bibfield  {author} {\bibinfo {author} {\bibfnamefont {A.}~\bibnamefont
  {Di~Piazza}}\ and\ \bibinfo {author} {\bibfnamefont {K.}~\bibnamefont {Qu}},\
  }\href@noop {} {\bibinfo {title} {Control of quantum radiation by squeezing
  the quantum vacuum fluctuation}} (\bibinfo {year} {2025}),\ \bibinfo {note}
  {to be submitted}\BibitemShut {NoStop}%
\bibitem [{\citenamefont {Franken}\ and\ \citenamefont
  {Ward}(1963)}]{Franken_RMP1963}%
  \BibitemOpen
  \bibfield  {author} {\bibinfo {author} {\bibfnamefont {P.~A.}\ \bibnamefont
  {Franken}}\ and\ \bibinfo {author} {\bibfnamefont {J.~F.}\ \bibnamefont
  {Ward}},\ }\bibfield  {title} {\bibinfo {title} {Optical harmonics and
  nonlinear phenomena},\ }\href {https://doi.org/10.1103/RevModPhys.35.23}
  {\bibfield  {journal} {\bibinfo  {journal} {Rev. Mod. Phys.}\ }\textbf
  {\bibinfo {volume} {35}},\ \bibinfo {pages} {23} (\bibinfo {year}
  {1963})}\BibitemShut {NoStop}%
\bibitem [{\citenamefont {Canagasabey}\ \emph {et~al.}(2009)\citenamefont
  {Canagasabey}, \citenamefont {Corbari}, \citenamefont {Gladyshev},
  \citenamefont {Liegeois}, \citenamefont {Guillemet}, \citenamefont
  {Hernandez}, \citenamefont {Yashkov}, \citenamefont {Kosolapov},
  \citenamefont {Dianov}, \citenamefont {Ibsen},\ and\ \citenamefont
  {Kazansky}}]{Canagasabey_09}%
  \BibitemOpen
  \bibfield  {author} {\bibinfo {author} {\bibfnamefont {A.}~\bibnamefont
  {Canagasabey}}, \bibinfo {author} {\bibfnamefont {C.}~\bibnamefont
  {Corbari}}, \bibinfo {author} {\bibfnamefont {A.~V.}\ \bibnamefont
  {Gladyshev}}, \bibinfo {author} {\bibfnamefont {F.}~\bibnamefont {Liegeois}},
  \bibinfo {author} {\bibfnamefont {S.}~\bibnamefont {Guillemet}}, \bibinfo
  {author} {\bibfnamefont {Y.}~\bibnamefont {Hernandez}}, \bibinfo {author}
  {\bibfnamefont {M.~V.}\ \bibnamefont {Yashkov}}, \bibinfo {author}
  {\bibfnamefont {A.}~\bibnamefont {Kosolapov}}, \bibinfo {author}
  {\bibfnamefont {E.~M.}\ \bibnamefont {Dianov}}, \bibinfo {author}
  {\bibfnamefont {M.}~\bibnamefont {Ibsen}},\ and\ \bibinfo {author}
  {\bibfnamefont {P.~G.}\ \bibnamefont {Kazansky}},\ }\bibfield  {title}
  {\bibinfo {title} {High-average-power second-harmonic generation from
  periodically poled silica fibers},\ }\href
  {https://doi.org/10.1364/OL.34.002483} {\bibfield  {journal} {\bibinfo
  {journal} {Opt. Lett.}\ }\textbf {\bibinfo {volume} {34}},\ \bibinfo {pages}
  {2483} (\bibinfo {year} {2009})}\BibitemShut {NoStop}%
\bibitem [{\citenamefont {Lin}\ \emph {et~al.}(2006)\citenamefont {Lin},
  \citenamefont {Yaman},\ and\ \citenamefont {Agrawal}}]{Lin_06fiber}%
  \BibitemOpen
  \bibfield  {author} {\bibinfo {author} {\bibfnamefont {Q.}~\bibnamefont
  {Lin}}, \bibinfo {author} {\bibfnamefont {F.}~\bibnamefont {Yaman}},\ and\
  \bibinfo {author} {\bibfnamefont {G.~P.}\ \bibnamefont {Agrawal}},\
  }\bibfield  {title} {\bibinfo {title} {Photon-pair generation by four-wave
  mixing in optical fibers},\ }\href {https://doi.org/10.1364/OL.31.001286}
  {\bibfield  {journal} {\bibinfo  {journal} {Opt. Lett.}\ }\textbf {\bibinfo
  {volume} {31}},\ \bibinfo {pages} {1286} (\bibinfo {year}
  {2006})}\BibitemShut {NoStop}%
\bibitem [{\citenamefont {Garay-Palmett}\ \emph {et~al.}(2023)\citenamefont
  {Garay-Palmett}, \citenamefont {Kim}, \citenamefont {Zhang}, \citenamefont
  {Dom\'{i}nguez-Serna}, \citenamefont {Lorenz},\ and\ \citenamefont
  {U'Ren}}]{Garay_Palmett:23}%
  \BibitemOpen
  \bibfield  {author} {\bibinfo {author} {\bibfnamefont {K.}~\bibnamefont
  {Garay-Palmett}}, \bibinfo {author} {\bibfnamefont {D.~B.}\ \bibnamefont
  {Kim}}, \bibinfo {author} {\bibfnamefont {Y.}~\bibnamefont {Zhang}}, \bibinfo
  {author} {\bibfnamefont {F.~A.}\ \bibnamefont {Dom\'{i}nguez-Serna}},
  \bibinfo {author} {\bibfnamefont {V.~O.}\ \bibnamefont {Lorenz}},\ and\
  \bibinfo {author} {\bibfnamefont {A.~B.}\ \bibnamefont {U'Ren}},\ }\bibfield
  {title} {\bibinfo {title} {Fiber-based photon-pair generation: tutorial},\
  }\href {https://doi.org/10.1364/JOSAB.478008} {\bibfield  {journal} {\bibinfo
   {journal} {J. Opt. Soc. Am. B}\ }\textbf {\bibinfo {volume} {40}},\ \bibinfo
  {pages} {469} (\bibinfo {year} {2023})}\BibitemShut {NoStop}%
\bibitem [{\citenamefont {Kim}\ and\ \citenamefont
  {Kumar}(1994)}]{Kim_prl1994}%
  \BibitemOpen
  \bibfield  {author} {\bibinfo {author} {\bibfnamefont {C.}~\bibnamefont
  {Kim}}\ and\ \bibinfo {author} {\bibfnamefont {P.}~\bibnamefont {Kumar}},\
  }\bibfield  {title} {\bibinfo {title} {Quadrature-squeezed light detection
  using a self-generated matched local oscillator},\ }\href
  {https://doi.org/10.1103/PhysRevLett.73.1605} {\bibfield  {journal} {\bibinfo
   {journal} {Phys. Rev. Lett.}\ }\textbf {\bibinfo {volume} {73}},\ \bibinfo
  {pages} {1605} (\bibinfo {year} {1994})}\BibitemShut {NoStop}%
\bibitem [{\citenamefont {Vahlbruch}\ \emph {et~al.}(2016)\citenamefont
  {Vahlbruch}, \citenamefont {Mehmet}, \citenamefont {Danzmann},\ and\
  \citenamefont {Schnabel}}]{PRL_15db_2016}%
  \BibitemOpen
  \bibfield  {author} {\bibinfo {author} {\bibfnamefont {H.}~\bibnamefont
  {Vahlbruch}}, \bibinfo {author} {\bibfnamefont {M.}~\bibnamefont {Mehmet}},
  \bibinfo {author} {\bibfnamefont {K.}~\bibnamefont {Danzmann}},\ and\
  \bibinfo {author} {\bibfnamefont {R.}~\bibnamefont {Schnabel}},\ }\bibfield
  {title} {\bibinfo {title} {Detection of 15 db squeezed states of light and
  their application for the absolute calibration of photoelectric quantum
  efficiency},\ }\href {https://doi.org/10.1103/PhysRevLett.117.110801}
  {\bibfield  {journal} {\bibinfo  {journal} {Phys. Rev. Lett.}\ }\textbf
  {\bibinfo {volume} {117}},\ \bibinfo {pages} {110801} (\bibinfo {year}
  {2016})}\BibitemShut {NoStop}%
\bibitem [{\citenamefont {Varma}\ \emph {et~al.}(2008)\citenamefont {Varma},
  \citenamefont {Chen},\ and\ \citenamefont {Milchberg}}]{Milchberg_prl2008}%
  \BibitemOpen
  \bibfield  {author} {\bibinfo {author} {\bibfnamefont {S.}~\bibnamefont
  {Varma}}, \bibinfo {author} {\bibfnamefont {Y.~H.}\ \bibnamefont {Chen}},\
  and\ \bibinfo {author} {\bibfnamefont {H.~M.}\ \bibnamefont {Milchberg}},\
  }\bibfield  {title} {\bibinfo {title} {{Trapping and Destruction of
  Long-Range High-Intensity Optical Filaments by Molecular Quantum Wakes in
  Air}},\ }\href {https://doi.org/10.1103/PhysRevLett.101.205001} {\bibfield
  {journal} {\bibinfo  {journal} {Phys. Rev. Lett.}\ }\textbf {\bibinfo
  {volume} {101}},\ \bibinfo {pages} {205001} (\bibinfo {year}
  {2008})}\BibitemShut {NoStop}%
\bibitem [{\citenamefont {Varma}\ \emph {et~al.}(2012)\citenamefont {Varma},
  \citenamefont {Chen}, \citenamefont {Palastro}, \citenamefont {Fallahkair},
  \citenamefont {Rosenthal}, \citenamefont {Antonsen},\ and\ \citenamefont
  {Milchberg}}]{Milchberg_pra2012}%
  \BibitemOpen
  \bibfield  {author} {\bibinfo {author} {\bibfnamefont {S.}~\bibnamefont
  {Varma}}, \bibinfo {author} {\bibfnamefont {Y.-H.}\ \bibnamefont {Chen}},
  \bibinfo {author} {\bibfnamefont {J.~P.}\ \bibnamefont {Palastro}}, \bibinfo
  {author} {\bibfnamefont {A.~B.}\ \bibnamefont {Fallahkair}}, \bibinfo
  {author} {\bibfnamefont {E.~W.}\ \bibnamefont {Rosenthal}}, \bibinfo {author}
  {\bibfnamefont {T.}~\bibnamefont {Antonsen}},\ and\ \bibinfo {author}
  {\bibfnamefont {H.~M.}\ \bibnamefont {Milchberg}},\ }\bibfield  {title}
  {\bibinfo {title} {Molecular quantum wake-induced pulse shaping and extension
  of femtosecond air filaments},\ }\href
  {https://doi.org/10.1103/PhysRevA.86.023850} {\bibfield  {journal} {\bibinfo
  {journal} {Phys. Rev. A}\ }\textbf {\bibinfo {volume} {86}},\ \bibinfo
  {pages} {023850} (\bibinfo {year} {2012})}\BibitemShut {NoStop}%
\bibitem [{\citenamefont {Zahedpour}\ \emph {et~al.}(2014)\citenamefont
  {Zahedpour}, \citenamefont {Wahlstrand},\ and\ \citenamefont
  {Milchberg}}]{Milchberg_prl2014}%
  \BibitemOpen
  \bibfield  {author} {\bibinfo {author} {\bibfnamefont {S.}~\bibnamefont
  {Zahedpour}}, \bibinfo {author} {\bibfnamefont {J.~K.}\ \bibnamefont
  {Wahlstrand}},\ and\ \bibinfo {author} {\bibfnamefont {H.~M.}\ \bibnamefont
  {Milchberg}},\ }\bibfield  {title} {\bibinfo {title} {{Quantum Control of
  Molecular Gas Hydrodynamics}},\ }\href
  {https://doi.org/10.1103/PhysRevLett.112.143601} {\bibfield  {journal}
  {\bibinfo  {journal} {Phys. Rev. Lett.}\ }\textbf {\bibinfo {volume} {112}},\
  \bibinfo {pages} {143601} (\bibinfo {year} {2014})}\BibitemShut {NoStop}%
\bibitem [{\citenamefont {Kominis}\ \emph {et~al.}(2014)\citenamefont
  {Kominis}, \citenamefont {Kolliopoulos}, \citenamefont {Charalambidis},\ and\
  \citenamefont {Tzallas}}]{PRA_Kominis2014}%
  \BibitemOpen
  \bibfield  {author} {\bibinfo {author} {\bibfnamefont {I.~K.}\ \bibnamefont
  {Kominis}}, \bibinfo {author} {\bibfnamefont {G.}~\bibnamefont
  {Kolliopoulos}}, \bibinfo {author} {\bibfnamefont {D.}~\bibnamefont
  {Charalambidis}},\ and\ \bibinfo {author} {\bibfnamefont {P.}~\bibnamefont
  {Tzallas}},\ }\bibfield  {title} {\bibinfo {title} {Quantum-optical nature of
  the recollision process in high-order-harmonic generation},\ }\href
  {https://doi.org/10.1103/PhysRevA.89.063827} {\bibfield  {journal} {\bibinfo
  {journal} {Phys. Rev. A}\ }\textbf {\bibinfo {volume} {89}},\ \bibinfo
  {pages} {063827} (\bibinfo {year} {2014})}\BibitemShut {NoStop}%
\bibitem [{\citenamefont {Gonoskov}\ \emph {et~al.}(2016)\citenamefont
  {Gonoskov}, \citenamefont {Tsatrafyllis}, \citenamefont {Kominis},\ and\
  \citenamefont {Tzallas}}]{sc_HHG2016}%
  \BibitemOpen
  \bibfield  {author} {\bibinfo {author} {\bibfnamefont {I.~A.}\ \bibnamefont
  {Gonoskov}}, \bibinfo {author} {\bibfnamefont {N.}~\bibnamefont
  {Tsatrafyllis}}, \bibinfo {author} {\bibfnamefont {I.~K.}\ \bibnamefont
  {Kominis}},\ and\ \bibinfo {author} {\bibfnamefont {P.}~\bibnamefont
  {Tzallas}},\ }\bibfield  {title} {\bibinfo {title} {Quantum optical
  signatures in strong-field laser physics: Infrared photon counting in
  high-order-harmonic generation},\ }\href {https://doi.org/10.1038/srep32821}
  {\bibfield  {journal} {\bibinfo  {journal} {Scientific Reports}\ }\textbf
  {\bibinfo {volume} {6}},\ \bibinfo {pages} {32821} (\bibinfo {year}
  {2016})}\BibitemShut {NoStop}%
\bibitem [{\citenamefont {Tsatrafyllis}\ \emph {et~al.}(2017)\citenamefont
  {Tsatrafyllis}, \citenamefont {Kominis}, \citenamefont {Gonoskov},\ and\
  \citenamefont {Tzallas}}]{nc_HHG2017}%
  \BibitemOpen
  \bibfield  {author} {\bibinfo {author} {\bibfnamefont {N.}~\bibnamefont
  {Tsatrafyllis}}, \bibinfo {author} {\bibfnamefont {I.~K.}\ \bibnamefont
  {Kominis}}, \bibinfo {author} {\bibfnamefont {I.~A.}\ \bibnamefont
  {Gonoskov}},\ and\ \bibinfo {author} {\bibfnamefont {P.}~\bibnamefont
  {Tzallas}},\ }\bibfield  {title} {\bibinfo {title} {High-order harmonics
  measured by the photon statistics of the infrared driving-field exiting the
  atomic medium},\ }\href {https://doi.org/10.1038/ncomms15170} {\bibfield
  {journal} {\bibinfo  {journal} {Nature Communications}\ }\textbf {\bibinfo
  {volume} {8}},\ \bibinfo {pages} {15170} (\bibinfo {year}
  {2017})}\BibitemShut {NoStop}%
\bibitem [{\citenamefont {Lewenstein}\ \emph {et~al.}(2021)\citenamefont
  {Lewenstein}, \citenamefont {Ciappina}, \citenamefont {Pisanty},
  \citenamefont {Rivera-Dean}, \citenamefont {Stammer}, \citenamefont
  {Lamprou},\ and\ \citenamefont {Tzallas}}]{np_Lewenstein2021}%
  \BibitemOpen
  \bibfield  {author} {\bibinfo {author} {\bibfnamefont {M.}~\bibnamefont
  {Lewenstein}}, \bibinfo {author} {\bibfnamefont {M.~F.}\ \bibnamefont
  {Ciappina}}, \bibinfo {author} {\bibfnamefont {E.}~\bibnamefont {Pisanty}},
  \bibinfo {author} {\bibfnamefont {J.}~\bibnamefont {Rivera-Dean}}, \bibinfo
  {author} {\bibfnamefont {P.}~\bibnamefont {Stammer}}, \bibinfo {author}
  {\bibfnamefont {T.}~\bibnamefont {Lamprou}},\ and\ \bibinfo {author}
  {\bibfnamefont {P.}~\bibnamefont {Tzallas}},\ }\bibfield  {title} {\bibinfo
  {title} {Generation of optical schr{\"o}dinger cat states in intense
  laser--matter interactions},\ }\href
  {https://doi.org/10.1038/s41567-021-01317-w} {\bibfield  {journal} {\bibinfo
  {journal} {Nature Physics}\ }\textbf {\bibinfo {volume} {17}},\ \bibinfo
  {pages} {1104} (\bibinfo {year} {2021})}\BibitemShut {NoStop}%
\bibitem [{\citenamefont {Tsatrafyllis}\ \emph {et~al.}(2019)\citenamefont
  {Tsatrafyllis}, \citenamefont {K\"uhn}, \citenamefont {Dumergue},
  \citenamefont {Foldi}, \citenamefont {Kahaly}, \citenamefont {Cormier},
  \citenamefont {Gonoskov}, \citenamefont {Kiss}, \citenamefont {Varju},
  \citenamefont {Varro},\ and\ \citenamefont {Tzallas}}]{PRL_Tsatrafyllis2019}%
  \BibitemOpen
  \bibfield  {author} {\bibinfo {author} {\bibfnamefont {N.}~\bibnamefont
  {Tsatrafyllis}}, \bibinfo {author} {\bibfnamefont {S.}~\bibnamefont
  {K\"uhn}}, \bibinfo {author} {\bibfnamefont {M.}~\bibnamefont {Dumergue}},
  \bibinfo {author} {\bibfnamefont {P.}~\bibnamefont {Foldi}}, \bibinfo
  {author} {\bibfnamefont {S.}~\bibnamefont {Kahaly}}, \bibinfo {author}
  {\bibfnamefont {E.}~\bibnamefont {Cormier}}, \bibinfo {author} {\bibfnamefont
  {I.~A.}\ \bibnamefont {Gonoskov}}, \bibinfo {author} {\bibfnamefont
  {B.}~\bibnamefont {Kiss}}, \bibinfo {author} {\bibfnamefont {K.}~\bibnamefont
  {Varju}}, \bibinfo {author} {\bibfnamefont {S.}~\bibnamefont {Varro}},\ and\
  \bibinfo {author} {\bibfnamefont {P.}~\bibnamefont {Tzallas}},\ }\bibfield
  {title} {\bibinfo {title} {Quantum optical signatures in a strong laser pulse
  after interaction with semiconductors},\ }\href
  {https://doi.org/10.1103/PhysRevLett.122.193602} {\bibfield  {journal}
  {\bibinfo  {journal} {Phys. Rev. Lett.}\ }\textbf {\bibinfo {volume} {122}},\
  \bibinfo {pages} {193602} (\bibinfo {year} {2019})}\BibitemShut {NoStop}%
\bibitem [{\citenamefont {Gonoskov}\ \emph {et~al.}(2024)\citenamefont
  {Gonoskov}, \citenamefont {Sondenheimer}, \citenamefont {H\"unecke},
  \citenamefont {Kartashov}, \citenamefont {Peschel},\ and\ \citenamefont
  {Gr\"afe}}]{GonoskovPRB2024}%
  \BibitemOpen
  \bibfield  {author} {\bibinfo {author} {\bibfnamefont {I.}~\bibnamefont
  {Gonoskov}}, \bibinfo {author} {\bibfnamefont {R.}~\bibnamefont
  {Sondenheimer}}, \bibinfo {author} {\bibfnamefont {C.}~\bibnamefont
  {H\"unecke}}, \bibinfo {author} {\bibfnamefont {D.}~\bibnamefont
  {Kartashov}}, \bibinfo {author} {\bibfnamefont {U.}~\bibnamefont {Peschel}},\
  and\ \bibinfo {author} {\bibfnamefont {S.}~\bibnamefont {Gr\"afe}},\
  }\bibfield  {title} {\bibinfo {title} {Nonclassical light generation and
  control from laser-driven semiconductor intraband excitations},\ }\href
  {https://doi.org/10.1103/PhysRevB.109.125110} {\bibfield  {journal} {\bibinfo
   {journal} {Phys. Rev. B}\ }\textbf {\bibinfo {volume} {109}},\ \bibinfo
  {pages} {125110} (\bibinfo {year} {2024})}\BibitemShut {NoStop}%
\bibitem [{\citenamefont {Tzur}\ \emph {et~al.}(2024)\citenamefont {Tzur},
  \citenamefont {Birk}, \citenamefont {Gorlach}, \citenamefont {Kaminer},
  \citenamefont {Kr\"uger},\ and\ \citenamefont {Cohen}}]{TzurPRR2024}%
  \BibitemOpen
  \bibfield  {author} {\bibinfo {author} {\bibfnamefont {M.~E.}\ \bibnamefont
  {Tzur}}, \bibinfo {author} {\bibfnamefont {M.}~\bibnamefont {Birk}}, \bibinfo
  {author} {\bibfnamefont {A.}~\bibnamefont {Gorlach}}, \bibinfo {author}
  {\bibfnamefont {I.}~\bibnamefont {Kaminer}}, \bibinfo {author} {\bibfnamefont
  {M.}~\bibnamefont {Kr\"uger}},\ and\ \bibinfo {author} {\bibfnamefont
  {O.}~\bibnamefont {Cohen}},\ }\bibfield  {title} {\bibinfo {title}
  {Generation of squeezed high-order harmonics},\ }\href
  {https://doi.org/10.1103/PhysRevResearch.6.033079} {\bibfield  {journal}
  {\bibinfo  {journal} {Phys. Rev. Res.}\ }\textbf {\bibinfo {volume} {6}},\
  \bibinfo {pages} {033079} (\bibinfo {year} {2024})}\BibitemShut {NoStop}%
\bibitem [{\citenamefont {Stammer}\ \emph {et~al.}(2022)\citenamefont
  {Stammer}, \citenamefont {Rivera-Dean}, \citenamefont {Lamprou},
  \citenamefont {Pisanty}, \citenamefont {Ciappina}, \citenamefont {Tzallas},\
  and\ \citenamefont {Lewenstein}}]{PRLStammer_2022}%
  \BibitemOpen
  \bibfield  {author} {\bibinfo {author} {\bibfnamefont {P.}~\bibnamefont
  {Stammer}}, \bibinfo {author} {\bibfnamefont {J.}~\bibnamefont
  {Rivera-Dean}}, \bibinfo {author} {\bibfnamefont {T.}~\bibnamefont
  {Lamprou}}, \bibinfo {author} {\bibfnamefont {E.}~\bibnamefont {Pisanty}},
  \bibinfo {author} {\bibfnamefont {M.~F.}\ \bibnamefont {Ciappina}}, \bibinfo
  {author} {\bibfnamefont {P.}~\bibnamefont {Tzallas}},\ and\ \bibinfo {author}
  {\bibfnamefont {M.}~\bibnamefont {Lewenstein}},\ }\bibfield  {title}
  {\bibinfo {title} {High photon number entangled states and coherent state
  superposition from the extreme ultraviolet to the far infrared},\ }\href
  {https://doi.org/10.1103/PhysRevLett.128.123603} {\bibfield  {journal}
  {\bibinfo  {journal} {Phys. Rev. Lett.}\ }\textbf {\bibinfo {volume} {128}},\
  \bibinfo {pages} {123603} (\bibinfo {year} {2022})}\BibitemShut {NoStop}%
\bibitem [{\citenamefont {Gorlach}\ \emph {et~al.}(2020)\citenamefont
  {Gorlach}, \citenamefont {Neufeld}, \citenamefont {Rivera}, \citenamefont
  {Cohen},\ and\ \citenamefont {Kaminer}}]{nc_HHG2020}%
  \BibitemOpen
  \bibfield  {author} {\bibinfo {author} {\bibfnamefont {A.}~\bibnamefont
  {Gorlach}}, \bibinfo {author} {\bibfnamefont {O.}~\bibnamefont {Neufeld}},
  \bibinfo {author} {\bibfnamefont {N.}~\bibnamefont {Rivera}}, \bibinfo
  {author} {\bibfnamefont {O.}~\bibnamefont {Cohen}},\ and\ \bibinfo {author}
  {\bibfnamefont {I.}~\bibnamefont {Kaminer}},\ }\bibfield  {title} {\bibinfo
  {title} {The quantum-optical nature of high harmonic generation},\ }\href
  {https://doi.org/10.1038/s41467-020-18218-w} {\bibfield  {journal} {\bibinfo
  {journal} {Nature Communications}\ }\textbf {\bibinfo {volume} {11}},\
  \bibinfo {pages} {4598} (\bibinfo {year} {2020})}\BibitemShut {NoStop}%
\bibitem [{\citenamefont {Stammer}\ \emph {et~al.}(2023)\citenamefont
  {Stammer}, \citenamefont {Rivera-Dean}, \citenamefont {Maxwell},
  \citenamefont {Lamprou}, \citenamefont {Ord\'o\~nez}, \citenamefont
  {Ciappina}, \citenamefont {Tzallas},\ and\ \citenamefont
  {Lewenstein}}]{PRXQuantum2023}%
  \BibitemOpen
  \bibfield  {author} {\bibinfo {author} {\bibfnamefont {P.}~\bibnamefont
  {Stammer}}, \bibinfo {author} {\bibfnamefont {J.}~\bibnamefont
  {Rivera-Dean}}, \bibinfo {author} {\bibfnamefont {A.}~\bibnamefont
  {Maxwell}}, \bibinfo {author} {\bibfnamefont {T.}~\bibnamefont {Lamprou}},
  \bibinfo {author} {\bibfnamefont {A.}~\bibnamefont {Ord\'o\~nez}}, \bibinfo
  {author} {\bibfnamefont {M.~F.}\ \bibnamefont {Ciappina}}, \bibinfo {author}
  {\bibfnamefont {P.}~\bibnamefont {Tzallas}},\ and\ \bibinfo {author}
  {\bibfnamefont {M.}~\bibnamefont {Lewenstein}},\ }\bibfield  {title}
  {\bibinfo {title} {Quantum electrodynamics of intense laser-matter
  interactions: A tool for quantum state engineering},\ }\href
  {https://doi.org/10.1103/PRXQuantum.4.010201} {\bibfield  {journal} {\bibinfo
   {journal} {PRX Quantum}\ }\textbf {\bibinfo {volume} {4}},\ \bibinfo {pages}
  {010201} (\bibinfo {year} {2023})}\BibitemShut {NoStop}%
\bibitem [{\citenamefont {Theidel}\ \emph {et~al.}(2024)\citenamefont
  {Theidel}, \citenamefont {Cotte}, \citenamefont {Sondenheimer}, \citenamefont
  {Shiriaeva}, \citenamefont {Froidevaux}, \citenamefont {Severin},
  \citenamefont {Merdji-Larue}, \citenamefont {Mosel}, \citenamefont
  {Fr\"ohlich}, \citenamefont {Weber}, \citenamefont {Morgner}, \citenamefont
  {Kovacev}, \citenamefont {Biegert},\ and\ \citenamefont
  {Merdji}}]{TheidelPRX2024}%
  \BibitemOpen
  \bibfield  {author} {\bibinfo {author} {\bibfnamefont {D.}~\bibnamefont
  {Theidel}}, \bibinfo {author} {\bibfnamefont {V.}~\bibnamefont {Cotte}},
  \bibinfo {author} {\bibfnamefont {R.}~\bibnamefont {Sondenheimer}}, \bibinfo
  {author} {\bibfnamefont {V.}~\bibnamefont {Shiriaeva}}, \bibinfo {author}
  {\bibfnamefont {M.}~\bibnamefont {Froidevaux}}, \bibinfo {author}
  {\bibfnamefont {V.}~\bibnamefont {Severin}}, \bibinfo {author} {\bibfnamefont
  {A.}~\bibnamefont {Merdji-Larue}}, \bibinfo {author} {\bibfnamefont
  {P.}~\bibnamefont {Mosel}}, \bibinfo {author} {\bibfnamefont
  {S.}~\bibnamefont {Fr\"ohlich}}, \bibinfo {author} {\bibfnamefont {K.-A.}\
  \bibnamefont {Weber}}, \bibinfo {author} {\bibfnamefont {U.}~\bibnamefont
  {Morgner}}, \bibinfo {author} {\bibfnamefont {M.}~\bibnamefont {Kovacev}},
  \bibinfo {author} {\bibfnamefont {J.}~\bibnamefont {Biegert}},\ and\ \bibinfo
  {author} {\bibfnamefont {H.}~\bibnamefont {Merdji}},\ }\bibfield  {title}
  {\bibinfo {title} {Evidence of the quantum optical nature of high-harmonic
  generation},\ }\href {https://doi.org/10.1103/PRXQuantum.5.040319} {\bibfield
   {journal} {\bibinfo  {journal} {PRX Quantum}\ }\textbf {\bibinfo {volume}
  {5}},\ \bibinfo {pages} {040319} (\bibinfo {year} {2024})}\BibitemShut
  {NoStop}%
\bibitem [{\citenamefont {Shvets}\ \emph {et~al.}(1998)\citenamefont {Shvets},
  \citenamefont {Fisch}, \citenamefont {Pukhov},\ and\ \citenamefont {Meyer-ter
  Vehn}}]{Shvets_1998}%
  \BibitemOpen
  \bibfield  {author} {\bibinfo {author} {\bibfnamefont {G.}~\bibnamefont
  {Shvets}}, \bibinfo {author} {\bibfnamefont {N.~J.}\ \bibnamefont {Fisch}},
  \bibinfo {author} {\bibfnamefont {A.}~\bibnamefont {Pukhov}},\ and\ \bibinfo
  {author} {\bibfnamefont {J.}~\bibnamefont {Meyer-ter Vehn}},\ }\bibfield
  {title} {\bibinfo {title} {Superradiant amplification of an ultrashort laser
  pulse in a plasma by a counterpropagating pump},\ }\href
  {https://doi.org/10.1103/PhysRevLett.81.4879} {\bibfield  {journal} {\bibinfo
   {journal} {Phys. Rev. Lett.}\ }\textbf {\bibinfo {volume} {81}},\ \bibinfo
  {pages} {4879} (\bibinfo {year} {1998})}\BibitemShut {NoStop}%
\bibitem [{\citenamefont {Malkin}\ \emph {et~al.}(1999)\citenamefont {Malkin},
  \citenamefont {Shvets},\ and\ \citenamefont {Fisch}}]{malkin99}%
  \BibitemOpen
  \bibfield  {author} {\bibinfo {author} {\bibfnamefont {V.~M.}\ \bibnamefont
  {Malkin}}, \bibinfo {author} {\bibfnamefont {G.}~\bibnamefont {Shvets}},\
  and\ \bibinfo {author} {\bibfnamefont {N.~J.}\ \bibnamefont {Fisch}},\
  }\bibfield  {title} {\bibinfo {title} {Fast compression of laser beams to
  highly overcritical powers},\ }\href@noop {} {\bibfield  {journal} {\bibinfo
  {journal} {Phys. Rev. Lett.}\ }\textbf {\bibinfo {volume} {82}},\ \bibinfo
  {pages} {4448} (\bibinfo {year} {1999})}\BibitemShut {NoStop}%
\bibitem [{\citenamefont {Cheng}\ \emph {et~al.}(2005)\citenamefont {Cheng},
  \citenamefont {Avitzour}, \citenamefont {Ping}, \citenamefont {Suckewer},
  \citenamefont {Fisch}, \citenamefont {Hur},\ and\ \citenamefont
  {Wurtele}}]{Cheng2005}%
  \BibitemOpen
  \bibfield  {author} {\bibinfo {author} {\bibfnamefont {W.}~\bibnamefont
  {Cheng}}, \bibinfo {author} {\bibfnamefont {Y.}~\bibnamefont {Avitzour}},
  \bibinfo {author} {\bibfnamefont {Y.}~\bibnamefont {Ping}}, \bibinfo {author}
  {\bibfnamefont {S.}~\bibnamefont {Suckewer}}, \bibinfo {author}
  {\bibfnamefont {N.~J.}\ \bibnamefont {Fisch}}, \bibinfo {author}
  {\bibfnamefont {M.~S.}\ \bibnamefont {Hur}},\ and\ \bibinfo {author}
  {\bibfnamefont {J.~S.}\ \bibnamefont {Wurtele}},\ }\bibfield  {title}
  {\bibinfo {title} {Reaching the nonlinear regime of raman amplification of
  ultrashort laser pulses},\ }\href
  {https://doi.org/10.1103/PhysRevLett.94.045003} {\bibfield  {journal}
  {\bibinfo  {journal} {Phys. Rev. Lett.}\ }\textbf {\bibinfo {volume} {94}},\
  \bibinfo {pages} {045003} (\bibinfo {year} {2005})}\BibitemShut {NoStop}%
\bibitem [{\citenamefont {Ren}\ \emph {et~al.}(2007)\citenamefont {Ren},
  \citenamefont {Cheng}, \citenamefont {Li},\ and\ \citenamefont
  {Suckewer}}]{Ren_np2007}%
  \BibitemOpen
  \bibfield  {author} {\bibinfo {author} {\bibfnamefont {J.}~\bibnamefont
  {Ren}}, \bibinfo {author} {\bibfnamefont {W.}~\bibnamefont {Cheng}}, \bibinfo
  {author} {\bibfnamefont {S.}~\bibnamefont {Li}},\ and\ \bibinfo {author}
  {\bibfnamefont {S.}~\bibnamefont {Suckewer}},\ }\bibfield  {title} {\bibinfo
  {title} {A new method for generating ultraintense and ultrashort laser
  pulses},\ }\href {https://doi.org/10.1038/nphys717} {\bibfield  {journal}
  {\bibinfo  {journal} {Nature Physics}\ }\textbf {\bibinfo {volume} {3}},\
  \bibinfo {pages} {732} (\bibinfo {year} {2007})}\BibitemShut {NoStop}%
\bibitem [{\citenamefont {Ren}\ \emph {et~al.}(2008)\citenamefont {Ren},
  \citenamefont {Li}, \citenamefont {Morozov}, \citenamefont {Suckewer},
  \citenamefont {Yampolsky}, \citenamefont {Malkin},\ and\ \citenamefont
  {Fisch}}]{Ren_PoP2008}%
  \BibitemOpen
  \bibfield  {author} {\bibinfo {author} {\bibfnamefont {J.}~\bibnamefont
  {Ren}}, \bibinfo {author} {\bibfnamefont {S.}~\bibnamefont {Li}}, \bibinfo
  {author} {\bibfnamefont {A.}~\bibnamefont {Morozov}}, \bibinfo {author}
  {\bibfnamefont {S.}~\bibnamefont {Suckewer}}, \bibinfo {author}
  {\bibfnamefont {N.~A.}\ \bibnamefont {Yampolsky}}, \bibinfo {author}
  {\bibfnamefont {V.~M.}\ \bibnamefont {Malkin}},\ and\ \bibinfo {author}
  {\bibfnamefont {N.~J.}\ \bibnamefont {Fisch}},\ }\bibfield  {title} {\bibinfo
  {title} {{A compact double-pass Raman backscattering
  amplifier/compressora}},\ }\href {https://doi.org/10.1063/1.2844352}
  {\bibfield  {journal} {\bibinfo  {journal} {Physics of Plasmas}\ }\textbf
  {\bibinfo {volume} {15}},\ \bibinfo {pages} {056702} (\bibinfo {year}
  {2008})}\BibitemShut {NoStop}%
\bibitem [{\citenamefont {Qu}\ \emph {et~al.}(2017)\citenamefont {Qu},
  \citenamefont {Barth},\ and\ \citenamefont {Fisch}}]{KQprl2017}%
  \BibitemOpen
  \bibfield  {author} {\bibinfo {author} {\bibfnamefont {K.}~\bibnamefont
  {Qu}}, \bibinfo {author} {\bibfnamefont {I.}~\bibnamefont {Barth}},\ and\
  \bibinfo {author} {\bibfnamefont {N.~J.}\ \bibnamefont {Fisch}},\ }\bibfield
  {title} {\bibinfo {title} {Plasma wave seed for raman amplifiers},\ }\href
  {https://doi.org/10.1103/PhysRevLett.118.164801} {\bibfield  {journal}
  {\bibinfo  {journal} {Phys. Rev. Lett.}\ }\textbf {\bibinfo {volume} {118}},\
  \bibinfo {pages} {164801} (\bibinfo {year} {2017})}\BibitemShut {NoStop}%
\bibitem [{\citenamefont {Vieux}\ \emph {et~al.}(2017)\citenamefont {Vieux},
  \citenamefont {Cipiccia}, \citenamefont {Grant}, \citenamefont {Lemos},
  \citenamefont {Grant}, \citenamefont {Ciocarlan}, \citenamefont {Ersfeld},
  \citenamefont {Hur}, \citenamefont {Lepipas}, \citenamefont {Manahan},
  \citenamefont {Raj}, \citenamefont {Reboredo~Gil}, \citenamefont {Subiel},
  \citenamefont {Welsh}, \citenamefont {Wiggins}, \citenamefont {Yoffe},
  \citenamefont {Farmer}, \citenamefont {Aniculaesei}, \citenamefont
  {Brunetti}, \citenamefont {Yang}, \citenamefont {Heathcote}, \citenamefont
  {Nersisyan}, \citenamefont {Lewis}, \citenamefont {Pukhov}, \citenamefont
  {Dias},\ and\ \citenamefont {Jaroszynski}}]{Vieux_Raman2017}%
  \BibitemOpen
  \bibfield  {author} {\bibinfo {author} {\bibfnamefont {G.}~\bibnamefont
  {Vieux}}, \bibinfo {author} {\bibfnamefont {S.}~\bibnamefont {Cipiccia}},
  \bibinfo {author} {\bibfnamefont {D.~W.}\ \bibnamefont {Grant}}, \bibinfo
  {author} {\bibfnamefont {N.}~\bibnamefont {Lemos}}, \bibinfo {author}
  {\bibfnamefont {P.}~\bibnamefont {Grant}}, \bibinfo {author} {\bibfnamefont
  {C.}~\bibnamefont {Ciocarlan}}, \bibinfo {author} {\bibfnamefont
  {B.}~\bibnamefont {Ersfeld}}, \bibinfo {author} {\bibfnamefont {M.~S.}\
  \bibnamefont {Hur}}, \bibinfo {author} {\bibfnamefont {P.}~\bibnamefont
  {Lepipas}}, \bibinfo {author} {\bibfnamefont {G.~G.}\ \bibnamefont
  {Manahan}}, \bibinfo {author} {\bibfnamefont {G.}~\bibnamefont {Raj}},
  \bibinfo {author} {\bibfnamefont {D.}~\bibnamefont {Reboredo~Gil}}, \bibinfo
  {author} {\bibfnamefont {A.}~\bibnamefont {Subiel}}, \bibinfo {author}
  {\bibfnamefont {G.~H.}\ \bibnamefont {Welsh}}, \bibinfo {author}
  {\bibfnamefont {S.~M.}\ \bibnamefont {Wiggins}}, \bibinfo {author}
  {\bibfnamefont {S.~R.}\ \bibnamefont {Yoffe}}, \bibinfo {author}
  {\bibfnamefont {J.~P.}\ \bibnamefont {Farmer}}, \bibinfo {author}
  {\bibfnamefont {C.}~\bibnamefont {Aniculaesei}}, \bibinfo {author}
  {\bibfnamefont {E.}~\bibnamefont {Brunetti}}, \bibinfo {author}
  {\bibfnamefont {X.}~\bibnamefont {Yang}}, \bibinfo {author} {\bibfnamefont
  {R.}~\bibnamefont {Heathcote}}, \bibinfo {author} {\bibfnamefont
  {G.}~\bibnamefont {Nersisyan}}, \bibinfo {author} {\bibfnamefont {C.~L.~S.}\
  \bibnamefont {Lewis}}, \bibinfo {author} {\bibfnamefont {A.}~\bibnamefont
  {Pukhov}}, \bibinfo {author} {\bibfnamefont {J.~M.}\ \bibnamefont {Dias}},\
  and\ \bibinfo {author} {\bibfnamefont {D.~A.}\ \bibnamefont {Jaroszynski}},\
  }\bibfield  {title} {\bibinfo {title} {An ultra-high gain and efficient
  amplifier based on raman amplification in plasma},\ }\href
  {https://doi.org/10.1038/s41598-017-01783-4} {\bibfield  {journal} {\bibinfo
  {journal} {Scientific Reports}\ }\textbf {\bibinfo {volume} {7}},\ \bibinfo
  {pages} {2399} (\bibinfo {year} {2017})}\BibitemShut {NoStop}%
\bibitem [{\citenamefont {Kirkwood}\ \emph {et~al.}(2018)\citenamefont
  {Kirkwood}, \citenamefont {Turnbull}, \citenamefont {Chapman}, \citenamefont
  {Wilks}, \citenamefont {Rosen}, \citenamefont {London}, \citenamefont
  {Pickworth}, \citenamefont {Dunlop}, \citenamefont {Moody}, \citenamefont
  {Strozzi}, \citenamefont {Michel}, \citenamefont {Divol}, \citenamefont
  {Landen}, \citenamefont {MacGowan}, \citenamefont {Van~Wonterghem},
  \citenamefont {Fournier},\ and\ \citenamefont {Blue}}]{np_Kirkwood2018}%
  \BibitemOpen
  \bibfield  {author} {\bibinfo {author} {\bibfnamefont {R.~K.}\ \bibnamefont
  {Kirkwood}}, \bibinfo {author} {\bibfnamefont {D.~P.}\ \bibnamefont
  {Turnbull}}, \bibinfo {author} {\bibfnamefont {T.}~\bibnamefont {Chapman}},
  \bibinfo {author} {\bibfnamefont {S.~C.}\ \bibnamefont {Wilks}}, \bibinfo
  {author} {\bibfnamefont {M.~D.}\ \bibnamefont {Rosen}}, \bibinfo {author}
  {\bibfnamefont {R.~A.}\ \bibnamefont {London}}, \bibinfo {author}
  {\bibfnamefont {L.~A.}\ \bibnamefont {Pickworth}}, \bibinfo {author}
  {\bibfnamefont {W.~H.}\ \bibnamefont {Dunlop}}, \bibinfo {author}
  {\bibfnamefont {J.~D.}\ \bibnamefont {Moody}}, \bibinfo {author}
  {\bibfnamefont {D.~J.}\ \bibnamefont {Strozzi}}, \bibinfo {author}
  {\bibfnamefont {P.~A.}\ \bibnamefont {Michel}}, \bibinfo {author}
  {\bibfnamefont {L.}~\bibnamefont {Divol}}, \bibinfo {author} {\bibfnamefont
  {O.~L.}\ \bibnamefont {Landen}}, \bibinfo {author} {\bibfnamefont {B.~J.}\
  \bibnamefont {MacGowan}}, \bibinfo {author} {\bibfnamefont {B.~M.}\
  \bibnamefont {Van~Wonterghem}}, \bibinfo {author} {\bibfnamefont {K.~B.}\
  \bibnamefont {Fournier}},\ and\ \bibinfo {author} {\bibfnamefont {B.~E.}\
  \bibnamefont {Blue}},\ }\bibfield  {title} {\bibinfo {title} {Plasma-based
  beam combiner for very high fluence and energy},\ }\href
  {https://doi.org/10.1038/nphys4271} {\bibfield  {journal} {\bibinfo
  {journal} {Nature Physics}\ }\textbf {\bibinfo {volume} {14}},\ \bibinfo
  {pages} {80} (\bibinfo {year} {2018})}\BibitemShut {NoStop}%
\bibitem [{\citenamefont {Qu}\ and\ \citenamefont
  {Fisch}(2024)}]{Qu_PRE_entangle24}%
  \BibitemOpen
  \bibfield  {author} {\bibinfo {author} {\bibfnamefont {K.}~\bibnamefont
  {Qu}}\ and\ \bibinfo {author} {\bibfnamefont {N.~J.}\ \bibnamefont {Fisch}},\
  }\bibfield  {title} {\bibinfo {title} {Producing entangled photon pairs and
  quantum squeezed states in plasmas},\ }\href
  {https://doi.org/10.1103/PhysRevE.110.065211} {\bibfield  {journal} {\bibinfo
   {journal} {Phys. Rev. E}\ }\textbf {\bibinfo {volume} {110}},\ \bibinfo
  {pages} {065211} (\bibinfo {year} {2024})}\BibitemShut {NoStop}%
\bibitem [{\citenamefont {Malkin}\ and\ \citenamefont
  {Fisch}(2020{\natexlab{a}})}]{Malkin_pre2020}%
  \BibitemOpen
  \bibfield  {author} {\bibinfo {author} {\bibfnamefont {V.~M.}\ \bibnamefont
  {Malkin}}\ and\ \bibinfo {author} {\bibfnamefont {N.~J.}\ \bibnamefont
  {Fisch}},\ }\bibfield  {title} {\bibinfo {title} {Towards megajoule x-ray
  lasers via relativistic four-photon cascade in plasma},\ }\href
  {https://doi.org/10.1103/PhysRevE.101.023211} {\bibfield  {journal} {\bibinfo
   {journal} {Phys. Rev. E}\ }\textbf {\bibinfo {volume} {101}},\ \bibinfo
  {pages} {023211} (\bibinfo {year} {2020}{\natexlab{a}})}\BibitemShut
  {NoStop}%
\bibitem [{\citenamefont {Malkin}\ and\ \citenamefont
  {Fisch}(2020{\natexlab{b}})}]{Malkin_pre2020_2}%
  \BibitemOpen
  \bibfield  {author} {\bibinfo {author} {\bibfnamefont {V.~M.}\ \bibnamefont
  {Malkin}}\ and\ \bibinfo {author} {\bibfnamefont {N.~J.}\ \bibnamefont
  {Fisch}},\ }\bibfield  {title} {\bibinfo {title} {Resonant four-photon
  scattering of collinear laser pulses in plasma},\ }\href
  {https://doi.org/10.1103/PhysRevE.102.063207} {\bibfield  {journal} {\bibinfo
   {journal} {Phys. Rev. E}\ }\textbf {\bibinfo {volume} {102}},\ \bibinfo
  {pages} {063207} (\bibinfo {year} {2020}{\natexlab{b}})}\BibitemShut
  {NoStop}%
\bibitem [{\citenamefont {Malkin}\ and\ \citenamefont
  {Fisch}(2022)}]{Malkin_pre2022}%
  \BibitemOpen
  \bibfield  {author} {\bibinfo {author} {\bibfnamefont {V.~M.}\ \bibnamefont
  {Malkin}}\ and\ \bibinfo {author} {\bibfnamefont {N.~J.}\ \bibnamefont
  {Fisch}},\ }\bibfield  {title} {\bibinfo {title} {Super-resonant four-photon
  collinear laser frequency multiplication in plasma},\ }\href
  {https://doi.org/10.1103/PhysRevE.105.045207} {\bibfield  {journal} {\bibinfo
   {journal} {Phys. Rev. E}\ }\textbf {\bibinfo {volume} {105}},\ \bibinfo
  {pages} {045207} (\bibinfo {year} {2022})}\BibitemShut {NoStop}%
\bibitem [{\citenamefont {Griffith}\ \emph {et~al.}(2021)\citenamefont
  {Griffith}, \citenamefont {Qu},\ and\ \citenamefont {Fisch}}]{Griffith2021}%
  \BibitemOpen
  \bibfield  {author} {\bibinfo {author} {\bibfnamefont {A.}~\bibnamefont
  {Griffith}}, \bibinfo {author} {\bibfnamefont {K.}~\bibnamefont {Qu}},\ and\
  \bibinfo {author} {\bibfnamefont {N.~J.}\ \bibnamefont {Fisch}},\ }\bibfield
  {title} {\bibinfo {title} {{Modulation-slippage trade-off in resonant
  four-wave upconversion}},\ }\href {https://doi.org/10.1063/5.0046695}
  {\bibfield  {journal} {\bibinfo  {journal} {Physics of Plasmas}\ }\textbf
  {\bibinfo {volume} {28}},\ \bibinfo {pages} {052112} (\bibinfo {year}
  {2021})}\BibitemShut {NoStop}%
\bibitem [{\citenamefont {Malkin}\ \emph {et~al.}(2000)\citenamefont {Malkin},
  \citenamefont {Shvets},\ and\ \citenamefont {Fisch}}]{Malkin2000}%
  \BibitemOpen
  \bibfield  {author} {\bibinfo {author} {\bibfnamefont {V.~M.}\ \bibnamefont
  {Malkin}}, \bibinfo {author} {\bibfnamefont {G.}~\bibnamefont {Shvets}},\
  and\ \bibinfo {author} {\bibfnamefont {N.~J.}\ \bibnamefont {Fisch}},\
  }\bibfield  {title} {\bibinfo {title} {{Detuned Raman Amplification of Short
  Laser Pulses in Plasma}},\ }\href
  {https://doi.org/10.1103/PhysRevLett.84.1208} {\bibfield  {journal} {\bibinfo
   {journal} {Phys. Rev. Lett.}\ }\textbf {\bibinfo {volume} {84}},\ \bibinfo
  {pages} {1208} (\bibinfo {year} {2000})}\BibitemShut {NoStop}%
\bibitem [{\citenamefont {Hur}\ \emph {et~al.}(2023)\citenamefont {Hur},
  \citenamefont {Ersfeld}, \citenamefont {Lee}, \citenamefont {Kim},
  \citenamefont {Roh}, \citenamefont {Lee}, \citenamefont {Song}, \citenamefont
  {Kumar}, \citenamefont {Yoffe}, \citenamefont {Jaroszynski},\ and\
  \citenamefont {Suk}}]{MinSup_np2023}%
  \BibitemOpen
  \bibfield  {author} {\bibinfo {author} {\bibfnamefont {M.~S.}\ \bibnamefont
  {Hur}}, \bibinfo {author} {\bibfnamefont {B.}~\bibnamefont {Ersfeld}},
  \bibinfo {author} {\bibfnamefont {H.}~\bibnamefont {Lee}}, \bibinfo {author}
  {\bibfnamefont {H.}~\bibnamefont {Kim}}, \bibinfo {author} {\bibfnamefont
  {K.}~\bibnamefont {Roh}}, \bibinfo {author} {\bibfnamefont {Y.}~\bibnamefont
  {Lee}}, \bibinfo {author} {\bibfnamefont {H.~S.}\ \bibnamefont {Song}},
  \bibinfo {author} {\bibfnamefont {M.}~\bibnamefont {Kumar}}, \bibinfo
  {author} {\bibfnamefont {S.}~\bibnamefont {Yoffe}}, \bibinfo {author}
  {\bibfnamefont {D.~A.}\ \bibnamefont {Jaroszynski}},\ and\ \bibinfo {author}
  {\bibfnamefont {H.}~\bibnamefont {Suk}},\ }\bibfield  {title} {\bibinfo
  {title} {Laser pulse compression by a density gradient plasma for exawatt to
  zettawatt lasers},\ }\href {https://doi.org/10.1038/s41566-023-01321-x}
  {\bibfield  {journal} {\bibinfo  {journal} {Nature Photonics}\ }\textbf
  {\bibinfo {volume} {17}},\ \bibinfo {pages} {1074} (\bibinfo {year}
  {2023})}\BibitemShut {NoStop}%
\bibitem [{\citenamefont {Braunstein}\ and\ \citenamefont {van
  Loock}(2005)}]{CV_rmp2005}%
  \BibitemOpen
  \bibfield  {author} {\bibinfo {author} {\bibfnamefont {S.~L.}\ \bibnamefont
  {Braunstein}}\ and\ \bibinfo {author} {\bibfnamefont {P.}~\bibnamefont {van
  Loock}},\ }\bibfield  {title} {\bibinfo {title} {Quantum information with
  continuous variables},\ }\href {https://doi.org/10.1103/RevModPhys.77.513}
  {\bibfield  {journal} {\bibinfo  {journal} {Rev. Mod. Phys.}\ }\textbf
  {\bibinfo {volume} {77}},\ \bibinfo {pages} {513} (\bibinfo {year}
  {2005})}\BibitemShut {NoStop}%
\bibitem [{\citenamefont {Weedbrook}\ \emph {et~al.}(2012)\citenamefont
  {Weedbrook}, \citenamefont {Pirandola}, \citenamefont {Garc\'{\i}a-Patr\'on},
  \citenamefont {Cerf}, \citenamefont {Ralph}, \citenamefont {Shapiro},\ and\
  \citenamefont {Lloyd}}]{GQI_rmp2012}%
  \BibitemOpen
  \bibfield  {author} {\bibinfo {author} {\bibfnamefont {C.}~\bibnamefont
  {Weedbrook}}, \bibinfo {author} {\bibfnamefont {S.}~\bibnamefont
  {Pirandola}}, \bibinfo {author} {\bibfnamefont {R.}~\bibnamefont
  {Garc\'{\i}a-Patr\'on}}, \bibinfo {author} {\bibfnamefont {N.~J.}\
  \bibnamefont {Cerf}}, \bibinfo {author} {\bibfnamefont {T.~C.}\ \bibnamefont
  {Ralph}}, \bibinfo {author} {\bibfnamefont {J.~H.}\ \bibnamefont {Shapiro}},\
  and\ \bibinfo {author} {\bibfnamefont {S.}~\bibnamefont {Lloyd}},\ }\bibfield
   {title} {\bibinfo {title} {Gaussian quantum information},\ }\href
  {https://doi.org/10.1103/RevModPhys.84.621} {\bibfield  {journal} {\bibinfo
  {journal} {Rev. Mod. Phys.}\ }\textbf {\bibinfo {volume} {84}},\ \bibinfo
  {pages} {621} (\bibinfo {year} {2012})}\BibitemShut {NoStop}%
\bibitem [{\citenamefont {Rabl}\ \emph {et~al.}(2004)\citenamefont {Rabl},
  \citenamefont {Shnirman},\ and\ \citenamefont {Zoller}}]{Rabl_prb2004}%
  \BibitemOpen
  \bibfield  {author} {\bibinfo {author} {\bibfnamefont {P.}~\bibnamefont
  {Rabl}}, \bibinfo {author} {\bibfnamefont {A.}~\bibnamefont {Shnirman}},\
  and\ \bibinfo {author} {\bibfnamefont {P.}~\bibnamefont {Zoller}},\
  }\bibfield  {title} {\bibinfo {title} {Generation of squeezed states of
  nanomechanical resonators by reservoir engineering},\ }\href
  {https://doi.org/10.1103/PhysRevB.70.205304} {\bibfield  {journal} {\bibinfo
  {journal} {Phys. Rev. B}\ }\textbf {\bibinfo {volume} {70}},\ \bibinfo
  {pages} {205304} (\bibinfo {year} {2004})}\BibitemShut {NoStop}%
\bibitem [{\citenamefont {Parkins}\ \emph {et~al.}(2006)\citenamefont
  {Parkins}, \citenamefont {Solano},\ and\ \citenamefont
  {Cirac}}]{Parkins_prl2006}%
  \BibitemOpen
  \bibfield  {author} {\bibinfo {author} {\bibfnamefont {A.~S.}\ \bibnamefont
  {Parkins}}, \bibinfo {author} {\bibfnamefont {E.}~\bibnamefont {Solano}},\
  and\ \bibinfo {author} {\bibfnamefont {J.~I.}\ \bibnamefont {Cirac}},\
  }\bibfield  {title} {\bibinfo {title} {Unconditional two-mode squeezing of
  separated atomic ensembles},\ }\href
  {https://doi.org/10.1103/PhysRevLett.96.053602} {\bibfield  {journal}
  {\bibinfo  {journal} {Phys. Rev. Lett.}\ }\textbf {\bibinfo {volume} {96}},\
  \bibinfo {pages} {053602} (\bibinfo {year} {2006})}\BibitemShut {NoStop}%
\bibitem [{\citenamefont {Purdy}\ \emph {et~al.}(2013)\citenamefont {Purdy},
  \citenamefont {Yu}, \citenamefont {Peterson}, \citenamefont {Kampel},\ and\
  \citenamefont {Regal}}]{Purdy_2013}%
  \BibitemOpen
  \bibfield  {author} {\bibinfo {author} {\bibfnamefont {T.~P.}\ \bibnamefont
  {Purdy}}, \bibinfo {author} {\bibfnamefont {P.-L.}\ \bibnamefont {Yu}},
  \bibinfo {author} {\bibfnamefont {R.~W.}\ \bibnamefont {Peterson}}, \bibinfo
  {author} {\bibfnamefont {N.~S.}\ \bibnamefont {Kampel}},\ and\ \bibinfo
  {author} {\bibfnamefont {C.~A.}\ \bibnamefont {Regal}},\ }\bibfield  {title}
  {\bibinfo {title} {Strong optomechanical squeezing of light},\ }\href
  {https://doi.org/10.1103/PhysRevX.3.031012} {\bibfield  {journal} {\bibinfo
  {journal} {Phys. Rev. X}\ }\textbf {\bibinfo {volume} {3}},\ \bibinfo {pages}
  {031012} (\bibinfo {year} {2013})}\BibitemShut {NoStop}%
\bibitem [{\citenamefont {Wang}\ and\ \citenamefont
  {Clerk}(2013)}]{Wangprl2013}%
  \BibitemOpen
  \bibfield  {author} {\bibinfo {author} {\bibfnamefont {Y.-D.}\ \bibnamefont
  {Wang}}\ and\ \bibinfo {author} {\bibfnamefont {A.~A.}\ \bibnamefont
  {Clerk}},\ }\bibfield  {title} {\bibinfo {title} {Reservoir-engineered
  entanglement in optomechanical systems},\ }\href
  {https://doi.org/10.1103/PhysRevLett.110.253601} {\bibfield  {journal}
  {\bibinfo  {journal} {Phys. Rev. Lett.}\ }\textbf {\bibinfo {volume} {110}},\
  \bibinfo {pages} {253601} (\bibinfo {year} {2013})}\BibitemShut {NoStop}%
\bibitem [{\citenamefont {Tian}(2013)}]{Tianprl2013}%
  \BibitemOpen
  \bibfield  {author} {\bibinfo {author} {\bibfnamefont {L.}~\bibnamefont
  {Tian}},\ }\bibfield  {title} {\bibinfo {title} {Robust photon entanglement
  via quantum interference in optomechanical interfaces},\ }\href
  {https://doi.org/10.1103/PhysRevLett.110.233602} {\bibfield  {journal}
  {\bibinfo  {journal} {Phys. Rev. Lett.}\ }\textbf {\bibinfo {volume} {110}},\
  \bibinfo {pages} {233602} (\bibinfo {year} {2013})}\BibitemShut {NoStop}%
\bibitem [{\citenamefont {Qu}\ and\ \citenamefont {Agarwal}(2014)}]{KQnjp2014}%
  \BibitemOpen
  \bibfield  {author} {\bibinfo {author} {\bibfnamefont {K.}~\bibnamefont
  {Qu}}\ and\ \bibinfo {author} {\bibfnamefont {G.~S.}\ \bibnamefont
  {Agarwal}},\ }\bibfield  {title} {\bibinfo {title} {Strong squeezing via
  phonon mediated spontaneous generation of photon pairs},\ }\href
  {https://doi.org/10.1088/1367-2630/16/11/113004} {\bibfield  {journal}
  {\bibinfo  {journal} {New Journal of Physics}\ }\textbf {\bibinfo {volume}
  {16}},\ \bibinfo {pages} {113004} (\bibinfo {year} {2014})}\BibitemShut
  {NoStop}%
\bibitem [{\citenamefont {Qu}\ and\ \citenamefont {Agarwal}(2015)}]{KQpra2015}%
  \BibitemOpen
  \bibfield  {author} {\bibinfo {author} {\bibfnamefont {K.}~\bibnamefont
  {Qu}}\ and\ \bibinfo {author} {\bibfnamefont {G.~S.}\ \bibnamefont
  {Agarwal}},\ }\bibfield  {title} {\bibinfo {title} {Generating quadrature
  squeezed light with dissipative optomechanical coupling},\ }\href
  {https://doi.org/10.1103/PhysRevA.91.063815} {\bibfield  {journal} {\bibinfo
  {journal} {Phys. Rev. A}\ }\textbf {\bibinfo {volume} {91}},\ \bibinfo
  {pages} {063815} (\bibinfo {year} {2015})}\BibitemShut {NoStop}%
\bibitem [{\citenamefont {Magrini}\ \emph {et~al.}(2022)\citenamefont
  {Magrini}, \citenamefont {Camarena-Ch\'avez}, \citenamefont {Bach},
  \citenamefont {Johnson},\ and\ \citenamefont {Aspelmeyer}}]{Magrini_2022}%
  \BibitemOpen
  \bibfield  {author} {\bibinfo {author} {\bibfnamefont {L.}~\bibnamefont
  {Magrini}}, \bibinfo {author} {\bibfnamefont {V.~A.}\ \bibnamefont
  {Camarena-Ch\'avez}}, \bibinfo {author} {\bibfnamefont {C.}~\bibnamefont
  {Bach}}, \bibinfo {author} {\bibfnamefont {A.}~\bibnamefont {Johnson}},\ and\
  \bibinfo {author} {\bibfnamefont {M.}~\bibnamefont {Aspelmeyer}},\ }\bibfield
   {title} {\bibinfo {title} {Squeezed light from a levitated nanoparticle at
  room temperature},\ }\href {https://doi.org/10.1103/PhysRevLett.129.053601}
  {\bibfield  {journal} {\bibinfo  {journal} {Phys. Rev. Lett.}\ }\textbf
  {\bibinfo {volume} {129}},\ \bibinfo {pages} {053601} (\bibinfo {year}
  {2022})}\BibitemShut {NoStop}%
\bibitem [{\citenamefont {Balakin}\ \emph {et~al.}(2003)\citenamefont
  {Balakin}, \citenamefont {Fraiman}, \citenamefont {Fisch},\ and\
  \citenamefont {Malkin}}]{Balakin2003}%
  \BibitemOpen
  \bibfield  {author} {\bibinfo {author} {\bibfnamefont {A.~A.}\ \bibnamefont
  {Balakin}}, \bibinfo {author} {\bibfnamefont {G.~M.}\ \bibnamefont
  {Fraiman}}, \bibinfo {author} {\bibfnamefont {N.~J.}\ \bibnamefont {Fisch}},\
  and\ \bibinfo {author} {\bibfnamefont {V.~M.}\ \bibnamefont {Malkin}},\
  }\bibfield  {title} {\bibinfo {title} {Noise suppression and enhanced
  focusability in plasma raman amplifier with multi-frequency pump},\ }\href
  {https://doi.org/10.1063/1.1621002} {\bibfield  {journal} {\bibinfo
  {journal} {Physics of Plasmas}\ }\textbf {\bibinfo {volume} {10}},\ \bibinfo
  {pages} {4856} (\bibinfo {year} {2003})}\BibitemShut {NoStop}%
\bibitem [{\citenamefont {Berger}\ \emph {et~al.}(2004)\citenamefont {Berger},
  \citenamefont {Clark}, \citenamefont {Solodov}, \citenamefont {Valeo},\ and\
  \citenamefont {Fisch}}]{Berger2004}%
  \BibitemOpen
  \bibfield  {author} {\bibinfo {author} {\bibfnamefont {R.~L.}\ \bibnamefont
  {Berger}}, \bibinfo {author} {\bibfnamefont {D.~S.}\ \bibnamefont {Clark}},
  \bibinfo {author} {\bibfnamefont {A.~A.}\ \bibnamefont {Solodov}}, \bibinfo
  {author} {\bibfnamefont {E.~J.}\ \bibnamefont {Valeo}},\ and\ \bibinfo
  {author} {\bibfnamefont {N.~J.}\ \bibnamefont {Fisch}},\ }\bibfield  {title}
  {\bibinfo {title} {Inverse bremsstrahlung stabilization of noise in the
  generation of ultrashort intense pulses by backward raman amplification},\
  }\href {https://doi.org/10.1063/1.1695356} {\bibfield  {journal} {\bibinfo
  {journal} {Physics of Plasmas}\ }\textbf {\bibinfo {volume} {11}},\ \bibinfo
  {pages} {1931} (\bibinfo {year} {2004})}\BibitemShut {NoStop}%
\bibitem [{\citenamefont {Sheffield}\ \emph {et~al.}(2010)\citenamefont
  {Sheffield}, \citenamefont {Froula}, \citenamefont {Glenzer},\ and\
  \citenamefont {Luhmann}}]{sheffield2010plasma}%
  \BibitemOpen
  \bibfield  {author} {\bibinfo {author} {\bibfnamefont {J.}~\bibnamefont
  {Sheffield}}, \bibinfo {author} {\bibfnamefont {D.}~\bibnamefont {Froula}},
  \bibinfo {author} {\bibfnamefont {S.}~\bibnamefont {Glenzer}},\ and\ \bibinfo
  {author} {\bibfnamefont {N.}~\bibnamefont {Luhmann}},\ }\href
  {https://books.google.com/books?id=1NS5Fxam1lkC} {\emph {\bibinfo {title}
  {Plasma Scattering of Electromagnetic Radiation: Theory and Measurement
  Techniques}}}\ (\bibinfo  {publisher} {Academic Press},\ \bibinfo {year}
  {2010})\BibitemShut {NoStop}%
\bibitem [{\citenamefont {C\'esar}\ \emph {et~al.}(2009)\citenamefont
  {C\'esar}, \citenamefont {Coelho}, \citenamefont {Cassemiro}, \citenamefont
  {Villar}, \citenamefont {Lassen}, \citenamefont {Nussenzveig},\ and\
  \citenamefont {Martinelli}}]{CrystalPhonon_pra2009}%
  \BibitemOpen
  \bibfield  {author} {\bibinfo {author} {\bibfnamefont {J.~E.~S.}\
  \bibnamefont {C\'esar}}, \bibinfo {author} {\bibfnamefont {A.~S.}\
  \bibnamefont {Coelho}}, \bibinfo {author} {\bibfnamefont {K.~N.}\
  \bibnamefont {Cassemiro}}, \bibinfo {author} {\bibfnamefont {A.~S.}\
  \bibnamefont {Villar}}, \bibinfo {author} {\bibfnamefont {M.}~\bibnamefont
  {Lassen}}, \bibinfo {author} {\bibfnamefont {P.}~\bibnamefont
  {Nussenzveig}},\ and\ \bibinfo {author} {\bibfnamefont {M.}~\bibnamefont
  {Martinelli}},\ }\bibfield  {title} {\bibinfo {title} {Extra phase noise from
  thermal fluctuations in nonlinear optical crystals},\ }\href
  {https://doi.org/10.1103/PhysRevA.79.063816} {\bibfield  {journal} {\bibinfo
  {journal} {Phys. Rev. A}\ }\textbf {\bibinfo {volume} {79}},\ \bibinfo
  {pages} {063816} (\bibinfo {year} {2009})}\BibitemShut {NoStop}%
\bibitem [{\citenamefont {Lax}(1966)}]{Lax_pr1966}%
  \BibitemOpen
  \bibfield  {author} {\bibinfo {author} {\bibfnamefont {M.}~\bibnamefont
  {Lax}},\ }\bibfield  {title} {\bibinfo {title} {Quantum noise. iv. quantum
  theory of noise sources},\ }\href {https://doi.org/10.1103/PhysRev.145.110}
  {\bibfield  {journal} {\bibinfo  {journal} {Phys. Rev.}\ }\textbf {\bibinfo
  {volume} {145}},\ \bibinfo {pages} {110} (\bibinfo {year}
  {1966})}\BibitemShut {NoStop}%
\bibitem [{\citenamefont {Umstadter}(2003)}]{Umstadter_2003}%
  \BibitemOpen
  \bibfield  {author} {\bibinfo {author} {\bibfnamefont {D.}~\bibnamefont
  {Umstadter}},\ }\bibfield  {title} {\bibinfo {title} {Relativistic
  laser–plasma interactions},\ }\href
  {https://doi.org/10.1088/0022-3727/36/8/202} {\bibfield  {journal} {\bibinfo
  {journal} {Journal of Physics D: Applied Physics}\ }\textbf {\bibinfo
  {volume} {36}},\ \bibinfo {pages} {R151} (\bibinfo {year}
  {2003})}\BibitemShut {NoStop}%
\bibitem [{\citenamefont {York}\ \emph {et~al.}(2008)\citenamefont {York},
  \citenamefont {Milchberg}, \citenamefont {Palastro},\ and\ \citenamefont
  {Antonsen}}]{York_prl2008}%
  \BibitemOpen
  \bibfield  {author} {\bibinfo {author} {\bibfnamefont {A.~G.}\ \bibnamefont
  {York}}, \bibinfo {author} {\bibfnamefont {H.~M.}\ \bibnamefont {Milchberg}},
  \bibinfo {author} {\bibfnamefont {J.~P.}\ \bibnamefont {Palastro}},\ and\
  \bibinfo {author} {\bibfnamefont {T.~M.}\ \bibnamefont {Antonsen}},\
  }\bibfield  {title} {\bibinfo {title} {Direct acceleration of electrons in a
  corrugated plasma waveguide},\ }\href
  {https://doi.org/10.1103/PhysRevLett.100.195001} {\bibfield  {journal}
  {\bibinfo  {journal} {Phys. Rev. Lett.}\ }\textbf {\bibinfo {volume} {100}},\
  \bibinfo {pages} {195001} (\bibinfo {year} {2008})}\BibitemShut {NoStop}%
\bibitem [{\citenamefont {Palastro}\ \emph {et~al.}(2025)\citenamefont
  {Palastro}, \citenamefont {Miller}, \citenamefont {Edwards}, \citenamefont
  {Elliott}, \citenamefont {Mack}, \citenamefont {Singh},\ and\ \citenamefont
  {Thomas}}]{palastro2025}%
  \BibitemOpen
  \bibfield  {author} {\bibinfo {author} {\bibfnamefont {J.~P.}\ \bibnamefont
  {Palastro}}, \bibinfo {author} {\bibfnamefont {K.~G.}\ \bibnamefont
  {Miller}}, \bibinfo {author} {\bibfnamefont {M.~R.}\ \bibnamefont {Edwards}},
  \bibinfo {author} {\bibfnamefont {A.~L.}\ \bibnamefont {Elliott}}, \bibinfo
  {author} {\bibfnamefont {L.~S.}\ \bibnamefont {Mack}}, \bibinfo {author}
  {\bibfnamefont {D.}~\bibnamefont {Singh}},\ and\ \bibinfo {author}
  {\bibfnamefont {A.~G.~R.}\ \bibnamefont {Thomas}},\ }\href@noop {} {\bibinfo
  {title} {Arbitrary-velocity laser pulses in plasma waveguides}} (\bibinfo
  {year} {2025}),\ \Eprint {https://arxiv.org/abs/2503.15690}
  {arXiv:2503.15690} \BibitemShut {NoStop}%
\bibitem [{\citenamefont {Marqu\`es}\ \emph {et~al.}(2019)\citenamefont
  {Marqu\`es}, \citenamefont {Lancia}, \citenamefont {Gangolf}, \citenamefont
  {Blecher}, \citenamefont {Bola\~nos}, \citenamefont {Fuchs}, \citenamefont
  {Willi}, \citenamefont {Amiranoff}, \citenamefont {Berger}, \citenamefont
  {Chiaramello}, \citenamefont {Weber},\ and\ \citenamefont
  {Riconda}}]{PBA_PRX2019}%
  \BibitemOpen
  \bibfield  {author} {\bibinfo {author} {\bibfnamefont {J.-R.}\ \bibnamefont
  {Marqu\`es}}, \bibinfo {author} {\bibfnamefont {L.}~\bibnamefont {Lancia}},
  \bibinfo {author} {\bibfnamefont {T.}~\bibnamefont {Gangolf}}, \bibinfo
  {author} {\bibfnamefont {M.}~\bibnamefont {Blecher}}, \bibinfo {author}
  {\bibfnamefont {S.}~\bibnamefont {Bola\~nos}}, \bibinfo {author}
  {\bibfnamefont {J.}~\bibnamefont {Fuchs}}, \bibinfo {author} {\bibfnamefont
  {O.}~\bibnamefont {Willi}}, \bibinfo {author} {\bibfnamefont
  {F.}~\bibnamefont {Amiranoff}}, \bibinfo {author} {\bibfnamefont {R.~L.}\
  \bibnamefont {Berger}}, \bibinfo {author} {\bibfnamefont {M.}~\bibnamefont
  {Chiaramello}}, \bibinfo {author} {\bibfnamefont {S.}~\bibnamefont {Weber}},\
  and\ \bibinfo {author} {\bibfnamefont {C.}~\bibnamefont {Riconda}},\
  }\bibfield  {title} {\bibinfo {title} {Joule-level high-efficiency energy
  transfer to subpicosecond laser pulses by a plasma-based amplifier},\ }\href
  {https://doi.org/10.1103/PhysRevX.9.021008} {\bibfield  {journal} {\bibinfo
  {journal} {Phys. Rev. X}\ }\textbf {\bibinfo {volume} {9}},\ \bibinfo {pages}
  {021008} (\bibinfo {year} {2019})}\BibitemShut {NoStop}%
\bibitem [{\citenamefont {Tanaka}\ \emph {et~al.}(1982)\citenamefont {Tanaka},
  \citenamefont {Goldman}, \citenamefont {Seka}, \citenamefont {Richardson},
  \citenamefont {Soures},\ and\ \citenamefont {Williams}}]{Tanaka_prl1982}%
  \BibitemOpen
  \bibfield  {author} {\bibinfo {author} {\bibfnamefont {K.}~\bibnamefont
  {Tanaka}}, \bibinfo {author} {\bibfnamefont {L.~M.}\ \bibnamefont {Goldman}},
  \bibinfo {author} {\bibfnamefont {W.}~\bibnamefont {Seka}}, \bibinfo {author}
  {\bibfnamefont {M.~C.}\ \bibnamefont {Richardson}}, \bibinfo {author}
  {\bibfnamefont {J.~M.}\ \bibnamefont {Soures}},\ and\ \bibinfo {author}
  {\bibfnamefont {E.~A.}\ \bibnamefont {Williams}},\ }\bibfield  {title}
  {\bibinfo {title} {{Stimulated Raman Scattering from uv-Laser-Produced
  Plasmas}},\ }\href {https://doi.org/10.1103/PhysRevLett.48.1179} {\bibfield
  {journal} {\bibinfo  {journal} {Phys. Rev. Lett.}\ }\textbf {\bibinfo
  {volume} {48}},\ \bibinfo {pages} {1179} (\bibinfo {year}
  {1982})}\BibitemShut {NoStop}%
\bibitem [{\citenamefont {Wilson}\ \emph {et~al.}(2012)\citenamefont {Wilson},
  \citenamefont {Tallents}, \citenamefont {Pasley}, \citenamefont {Whittaker},
  \citenamefont {Rose}, \citenamefont {Guilbaud}, \citenamefont {Cassou},
  \citenamefont {Kazamias}, \citenamefont {Daboussi}, \citenamefont {Pittman},
  \citenamefont {Delmas}, \citenamefont {Demailly}, \citenamefont {Neveu},\
  and\ \citenamefont {Ros}}]{Wilson_pre2012}%
  \BibitemOpen
  \bibfield  {author} {\bibinfo {author} {\bibfnamefont {L.~A.}\ \bibnamefont
  {Wilson}}, \bibinfo {author} {\bibfnamefont {G.~J.}\ \bibnamefont
  {Tallents}}, \bibinfo {author} {\bibfnamefont {J.}~\bibnamefont {Pasley}},
  \bibinfo {author} {\bibfnamefont {D.~S.}\ \bibnamefont {Whittaker}}, \bibinfo
  {author} {\bibfnamefont {S.~J.}\ \bibnamefont {Rose}}, \bibinfo {author}
  {\bibfnamefont {O.}~\bibnamefont {Guilbaud}}, \bibinfo {author}
  {\bibfnamefont {K.}~\bibnamefont {Cassou}}, \bibinfo {author} {\bibfnamefont
  {S.}~\bibnamefont {Kazamias}}, \bibinfo {author} {\bibfnamefont
  {S.}~\bibnamefont {Daboussi}}, \bibinfo {author} {\bibfnamefont
  {M.}~\bibnamefont {Pittman}}, \bibinfo {author} {\bibfnamefont
  {O.}~\bibnamefont {Delmas}}, \bibinfo {author} {\bibfnamefont
  {J.}~\bibnamefont {Demailly}}, \bibinfo {author} {\bibfnamefont
  {O.}~\bibnamefont {Neveu}},\ and\ \bibinfo {author} {\bibfnamefont
  {D.}~\bibnamefont {Ros}},\ }\bibfield  {title} {\bibinfo {title} {Energy
  transport in short-pulse-laser-heated targets measured using extreme
  ultraviolet laser backlighting},\ }\href
  {https://doi.org/10.1103/PhysRevE.86.026406} {\bibfield  {journal} {\bibinfo
  {journal} {Phys. Rev. E}\ }\textbf {\bibinfo {volume} {86}},\ \bibinfo
  {pages} {026406} (\bibinfo {year} {2012})}\BibitemShut {NoStop}%
\bibitem [{\citenamefont {Manes}\ \emph {et~al.}(1985)\citenamefont {Manes},
  \citenamefont {Barr}, \citenamefont {Bliss}, \citenamefont {Drake},
  \citenamefont {Godwin}, \citenamefont {Gritton}, \citenamefont {Hildum},
  \citenamefont {Holloway}, \citenamefont {Hurley}, \citenamefont {Johnson},\
  and\ \citenamefont {et~al.}}]{Manes_1985}%
  \BibitemOpen
  \bibfield  {author} {\bibinfo {author} {\bibfnamefont {K.~R.}\ \bibnamefont
  {Manes}}, \bibinfo {author} {\bibfnamefont {O.~C.}\ \bibnamefont {Barr}},
  \bibinfo {author} {\bibfnamefont {E.~S.}\ \bibnamefont {Bliss}}, \bibinfo
  {author} {\bibfnamefont {R.~P.}\ \bibnamefont {Drake}}, \bibinfo {author}
  {\bibfnamefont {R.~O.}\ \bibnamefont {Godwin}}, \bibinfo {author}
  {\bibfnamefont {D.~G.}\ \bibnamefont {Gritton}}, \bibinfo {author}
  {\bibfnamefont {J.~S.}\ \bibnamefont {Hildum}}, \bibinfo {author}
  {\bibfnamefont {F.~W.}\ \bibnamefont {Holloway}}, \bibinfo {author}
  {\bibfnamefont {C.~A.}\ \bibnamefont {Hurley}}, \bibinfo {author}
  {\bibfnamefont {B.~C.}\ \bibnamefont {Johnson}},\ and\ \bibinfo {author}
  {\bibnamefont {et~al.}},\ }\bibfield  {title} {\bibinfo {title} {Novette
  facility: activation and experimental results},\ }\href
  {https://doi.org/10.1017/S0263034600001373} {\bibfield  {journal} {\bibinfo
  {journal} {Laser and Particle Beams}\ }\textbf {\bibinfo {volume} {3}},\
  \bibinfo {pages} {173–188} (\bibinfo {year} {1985})}\BibitemShut {NoStop}%
\bibitem [{\citenamefont {Drake}\ \emph {et~al.}(1989)\citenamefont {Drake},
  \citenamefont {Turner}, \citenamefont {Lasinski}, \citenamefont {Williams},
  \citenamefont {Estabrook}, \citenamefont {Kruer}, \citenamefont {Campbell},\
  and\ \citenamefont {Johnston}}]{Drake_pra1989}%
  \BibitemOpen
  \bibfield  {author} {\bibinfo {author} {\bibfnamefont {R.~P.}\ \bibnamefont
  {Drake}}, \bibinfo {author} {\bibfnamefont {R.~E.}\ \bibnamefont {Turner}},
  \bibinfo {author} {\bibfnamefont {B.~F.}\ \bibnamefont {Lasinski}}, \bibinfo
  {author} {\bibfnamefont {E.~A.}\ \bibnamefont {Williams}}, \bibinfo {author}
  {\bibfnamefont {K.}~\bibnamefont {Estabrook}}, \bibinfo {author}
  {\bibfnamefont {W.~L.}\ \bibnamefont {Kruer}}, \bibinfo {author}
  {\bibfnamefont {E.~M.}\ \bibnamefont {Campbell}},\ and\ \bibinfo {author}
  {\bibfnamefont {T.~W.}\ \bibnamefont {Johnston}},\ }\bibfield  {title}
  {\bibinfo {title} {X-ray emission caused by raman scattering in
  long-scale-length plasmas},\ }\href
  {https://doi.org/10.1103/PhysRevA.40.3219} {\bibfield  {journal} {\bibinfo
  {journal} {Phys. Rev. A}\ }\textbf {\bibinfo {volume} {40}},\ \bibinfo
  {pages} {3219} (\bibinfo {year} {1989})}\BibitemShut {NoStop}%
\bibitem [{\citenamefont {Glenzer}\ \emph {et~al.}(2003)\citenamefont
  {Glenzer}, \citenamefont {Gregori}, \citenamefont {Lee}, \citenamefont
  {Rogers}, \citenamefont {Pollaine},\ and\ \citenamefont
  {Landen}}]{Glenzer_prl2003}%
  \BibitemOpen
  \bibfield  {author} {\bibinfo {author} {\bibfnamefont {S.~H.}\ \bibnamefont
  {Glenzer}}, \bibinfo {author} {\bibfnamefont {G.}~\bibnamefont {Gregori}},
  \bibinfo {author} {\bibfnamefont {R.~W.}\ \bibnamefont {Lee}}, \bibinfo
  {author} {\bibfnamefont {F.~J.}\ \bibnamefont {Rogers}}, \bibinfo {author}
  {\bibfnamefont {S.~W.}\ \bibnamefont {Pollaine}},\ and\ \bibinfo {author}
  {\bibfnamefont {O.~L.}\ \bibnamefont {Landen}},\ }\bibfield  {title}
  {\bibinfo {title} {{Demonstration of Spectrally Resolved X-Ray Scattering in
  Dense Plasmas}},\ }\href {https://doi.org/10.1103/PhysRevLett.90.175002}
  {\bibfield  {journal} {\bibinfo  {journal} {Phys. Rev. Lett.}\ }\textbf
  {\bibinfo {volume} {90}},\ \bibinfo {pages} {175002} (\bibinfo {year}
  {2003})}\BibitemShut {NoStop}%
\bibitem [{\citenamefont {Glenzer}\ \emph {et~al.}(2007)\citenamefont
  {Glenzer}, \citenamefont {Landen}, \citenamefont {Neumayer}, \citenamefont
  {Lee}, \citenamefont {Widmann}, \citenamefont {Pollaine}, \citenamefont
  {Wallace}, \citenamefont {Gregori}, \citenamefont {H\"oll}, \citenamefont
  {Bornath}, \citenamefont {Thiele}, \citenamefont {Schwarz}, \citenamefont
  {Kraeft},\ and\ \citenamefont {Redmer}}]{Glenzer_prl2007}%
  \BibitemOpen
  \bibfield  {author} {\bibinfo {author} {\bibfnamefont {S.~H.}\ \bibnamefont
  {Glenzer}}, \bibinfo {author} {\bibfnamefont {O.~L.}\ \bibnamefont {Landen}},
  \bibinfo {author} {\bibfnamefont {P.}~\bibnamefont {Neumayer}}, \bibinfo
  {author} {\bibfnamefont {R.~W.}\ \bibnamefont {Lee}}, \bibinfo {author}
  {\bibfnamefont {K.}~\bibnamefont {Widmann}}, \bibinfo {author} {\bibfnamefont
  {S.~W.}\ \bibnamefont {Pollaine}}, \bibinfo {author} {\bibfnamefont {R.~J.}\
  \bibnamefont {Wallace}}, \bibinfo {author} {\bibfnamefont {G.}~\bibnamefont
  {Gregori}}, \bibinfo {author} {\bibfnamefont {A.}~\bibnamefont {H\"oll}},
  \bibinfo {author} {\bibfnamefont {T.}~\bibnamefont {Bornath}}, \bibinfo
  {author} {\bibfnamefont {R.}~\bibnamefont {Thiele}}, \bibinfo {author}
  {\bibfnamefont {V.}~\bibnamefont {Schwarz}}, \bibinfo {author} {\bibfnamefont
  {W.-D.}\ \bibnamefont {Kraeft}},\ and\ \bibinfo {author} {\bibfnamefont
  {R.}~\bibnamefont {Redmer}},\ }\bibfield  {title} {\bibinfo {title}
  {{Observations of Plasmons in Warm Dense Matter}},\ }\href
  {https://doi.org/10.1103/PhysRevLett.98.065002} {\bibfield  {journal}
  {\bibinfo  {journal} {Phys. Rev. Lett.}\ }\textbf {\bibinfo {volume} {98}},\
  \bibinfo {pages} {065002} (\bibinfo {year} {2007})}\BibitemShut {NoStop}%
\end{thebibliography}%

\end{document}